%% file: final_version.tex
\documentclass[fleqn,usenatbib]{mnras}

\input{additional_tex/pack}        
\input{additional_tex/definitions} 

\usepackage{newtxtext,newtxmath}

\usepackage[T1]{fontenc}

\DeclareRobustCommand{\VAN}[3]{#2}
\let\VANthebibliography\thebibliography
\def\thebibliography{\DeclareRobustCommand{\VAN}[3]{##3}\VANthebibliography}

\usepackage{graphicx}	
\usepackage{amsmath}	

\defcitealias{pizzati2023}{P24}
\def\citeP24{\citetalias{pizzati2023}}

\defcitealias{eilers2024}{E24}
\def\citeE24{\citetalias{eilers2024}}

\title[Modeling quasar-galaxy clustering at $z\approx6$]{A unified model for the clustering of quasars and galaxies at $z\approx6$}

\author[Pizzati et al.]{Elia Pizzati$^{1}$\thanks{\href{mailto:pizzati@strw.leidenuniv.nl}{pizzati@strw.leidenuniv.nl}},
Joseph F. Hennawi$^{1,2}$,
Joop Schaye$^{1}$, 
Matthieu Schaller$^{3,1}$,
Anna-Christina Eilers$^{4}$,
\newauthor
Feige Wang$^{5}$,
Carlos S.\ Frenk$^{6}$,
Willem Elbers$^{6}$,
John C.\ Helly$^{6}$,
Ruari Mackenzie$^{7}$,
Jorryt Matthee$^{8}$,
\newauthor
Rongmon Bordoloi$^{9}$,
Daichi Kashino$^{10}$,
Rohan P.\ Naidu$^{4}$\thanks{NHFP Hubble Fellow},
Minghao Yue$^{4}$
\\
$^{1}$ Leiden Observatory, Leiden University, P.O. Box 9513, 2300 RA Leiden,
The Netherlands\\
$^{2}$ Department of Physics, University of California, Santa Barbara, CA 93106, USA\\
$^{3}$ Lorentz Institute for Theoretical Physics, Leiden University, PO Box 9506, NL-2300 RA Leiden, The Netherlands\\
$^{4}$ MIT Kavli Institute for Astrophysics and Space Research, Massachusetts Institute of Technology, Cambridge, MA 02139, USA\\
$^{5}$ Steward Observatory, University of Arizona, 933 N Cherry Avenue, Tucson, AZ 85721, USA\\
$^{6}$ Institute for Computational Cosmology, Department of Physics, University of Durham, South Road, Durham, DH1 3LE, UK\\
$^{7}$ Department of Physics, ETH Z{\"u}rich, Wolfgang-Pauli-Strasse 27, Z{\"u}rich, 8093, Switzerland\\
$^{8}$ Institute of Science and Technology Austria (ISTA), Am Campus 1, 3400 Klosterneuburg, Austria\\
$^{9}$ Department of Physics, North Carolina State University, Raleigh, 27695, North Carolina, USA\\
$^{10}$ National Astronomical Observatory of Japan, 2-21-1 Osawa, Mitaka, Tokyo 181-8588, Japan
}

\date{Accepted XXX. Received YYY; in original form ZZZ}

\pubyear{2023}

\begin{document}
\label{firstpage}
\pagerange{\pageref{firstpage}--\pageref{lastpage}}
\maketitle

\begin{abstract}
Recent observations from the EIGER JWST program have measured for the first time the quasar-galaxy cross-correlation function at $z\approx6$. The 
auto-correlation function of faint $z\approx6$ quasars was also recently estimated. 
These measurements provide key insights into the properties of quasars and galaxies at high redshift and their relation with the host dark matter halos. 
In this work, we interpret these data building upon 
an empirical quasar population model that has been applied successfully to quasar clustering and demographic measurements at $z\approx2-4$. We make use of a new, large-volume N-body simulation with more than a trillion particles,
FLAMINGO-10k, to model quasars and galaxies simultaneously.
We successfully reproduce observations of $z\approx6$ quasars and galaxies (i.e., their clustering properties and luminosity functions), and infer key quantities such as their luminosity-halo mass relation, the mass function of their host halos, and their duty cycle/occupation fraction. Our key findings are: (i) quasars reside on average in $\approx10^{12.5}\,\msun$ halos (corresponding to $\approx5\sigma$ fluctuations in the initial conditions of the linear density field), 
but the distribution of host halo masses is quite broad; (ii) the duty cycle of (UV-bright) quasar activity is relatively low ($\approx1\%$); (iii) galaxies (that are bright in \hbox{[O~$\scriptstyle\rm III $]}) live in much smaller halos ($\approx10^{10.9}\,\msun$) and have a larger duty cycle (occupation fraction) of
$\approx13\%$. Finally, we focus on the inferred properties of quasars and present a homogeneous analysis of their evolution with redshift. The picture that emerges reveals a strong evolution of the host halo mass and duty cycle of quasars at $z\approx2-6$, and calls for new investigations of the role of quasar activity across cosmic time. 
\end{abstract}

\begin{keywords}
 large-scale structure of Universe -- quasars: general -- quasars: supermassive black holes -- galaxies: high-redshift
\end{keywords}



\section{Introduction}\label{sec:introduction}

Supermassive black holes (SMBHs) are thought to be ubiquitous in the Universe, residing at the center of almost every massive galaxy \citep[e.g.,][]{magorrian_1998, ferrarese_merritt2000, kormendy_ho2013}. The basic elements of our formation story for these enigmatic objects have hardly changed since their existence was hypothesized, triggered by the discovery of the first quasar \citep[][]{schmidt_1963}. Luminous quasars are powered by accretion onto a SMBH \citep{salpeter1964, zeldovich_1964, lynden_bell1969} and the
rest mass energy of this material is divided between the small fraction ($\approx 10\%$) of radiation that we observe, and the growth of the black hole. This implies that the growth of black holes is directly related to the accretion of material powering bright quasars. 

But this half-century-old picture is challenged by the existence of luminous high-$z$ quasars powered by $\gtrsim10^9\,\msun$ SMBHs at $z\gtrsim6$, well into the epoch of reionization \citep[EoR;][]{Mazzucchelli2017, Farina2023, fan2022}. Even more puzzling, quasars with similar masses have been discovered at $z\approx7.5$, merely 700 Myr after the Big Bang \citep{Banados_2018, Yang_2020a,yang2021, Wang_2021}. The advent of the James Webb Space Telescope (JWST) has made these findings even more compelling, 
with the record-breaking discoveries of moderately massive SMBHs ($\approx10^6-10^8\,\msun$) at even higher redshift ($z\approx8-11$; e.g., \citealt{Ubler2023, Maiolino2023, Kokorev2023, Larson2023, Bogdan2023}). 
How these SMBHs have formed at such early times challenges our understanding of black hole formation and growth.  
There does not appear to be enough cosmic time to grow them from the $100\,\msun$ seed black holes expected for Pop III stellar remnants \citep{Heger03}, even if they accrete at the maximal Eddington rate. This has led to an industry of speculation that SMBHs formed from far more massive seeds forming via direct collapse \citep[e.g.,][]{BL03} or coalescence of a dense Pop III star cluster \citep[e.g.,][]{Omukai08}. 

Addressing this challenge requires integrating SMBH growth into our current picture of galaxy formation and evolution. The tight
local scaling relation between SMBHs and galaxy bulges \citep{magorrian_1998}, as well as the need to tap into SMBH accretion as a source of energetic feedback 
that regulates star formation in massive galaxies \citep[e.g.,][]{benson2003,springel2005,bower2006}, has led to the modern picture that SMBHs and their host galaxies co-evolve \citep[][]{bower2017}{}{}.  
In this
context, an assortment of cosmological simulation models can produce the massive SMBHs \citep[e.g.,][]{feng15,MassiveBlack}
that are powering bright high-$z$ quasars 
starting with massive $\gtrsim 10^4~\msun$ seed black holes. These models generically predict that such quasars are hosted by massive ($M_\star \gtrsim 10^{11}~\msun$)
and highly star-forming (${\rm SFR} \gtrsim 100~\msun~{\rm yr}^{-1}$) galaxies, and reside in the rarest $M \gtrsim
10^{12.5}~\msun$ halos situated in the most overdense regions of the Universe \citep{DiMatteo2012,Costa14,feng15,MassiveBlack, Barai2018, Valentini2021}. 
While these numerical studies
establish the plausibility of the existence of high-$z$ quasars, rigorous tests of this theoretical picture have been lacking \citep[][]{fan2022, Habouzit2019}{}{}.

The key to understanding high-$z$ quasars and SMBH formation in a cosmological context is determining how they are embedded in the evolving cosmic web of dark matter (DM) halos that forms the backbone of all structures in the Universe according to the hierarchical structure formation paradigm.  ${\rm \Lambda}$CDM dictates that the clustering of a population of objects, or equivalently the size of the cosmic over-densities that they reside in, is directly related to their host halo masses \citep[e.g.][]{kaiser1984, bardeen1986, mo_white1996}. Measuring the masses of the halos that host bright quasars gives precious information not only on the large-scale environment that quasars inhabit, but also -- by comparing the observed abundance of quasars with that of the hosting halos -- on the fraction of SMBHs that are active as bright quasars at any given time (i.e., the {\it quasar duty cycle}). In turn, this fraction can be related to the total time SMBHs shine as quasars (or {\it quasar lifetime}, $t_\mathrm{Q}$; see e.g., \citealt{martini2001, haiman_hui2001, martini_2004}), which is an essential quantity for determining the growth of SMBHs and sets an upper limit to the characteristic timescale of quasar events. 
For these reasons, a measurement of the clustering of quasars at high redshift is key to unraveling their formation history \citep[e.g.,][]{cole_kaiser1989, efstathiou_rees1998}{}{}.

Quasar clustering studies at lower redshifts are already a fundamental ingredient on which we built our understanding of SMBHs, their accretion mechanisms, and the co-evolution with their host galaxies. Large-sky surveys, such as the Sloan Digital Sky Survey \citep[SDSS,][]{york2000} and the 2dF QSO redshift survey \citep[2QZ,][]{croom2004}, have delivered measurements of the auto-correlation function of quasars up to $z\approx4$ \citep[][]{porciani_2004,croom2005,porciani_norberg2006,shen2007,ross2009,eftekharzadeh2015}{}{}. These measurements reveal that in the last ten billion years ($z\lesssim2$), quasars have been tracing halos in a way that is similar to optically selected galaxies, with a linear bias factor close to unity \citep[][]{croom2005, ross2009}. This implies that quasars are hosted, on average, by common, $\approx10^{12}\,\msun$ halos which, incidentally, are also the ones with the highest star formation efficiency \citep[e.g.,][]{eke2004,fanidakis2010,fanidakis2013}{}{}. At $z\approx2-4$, however, the clustering of quasars shows a dramatic change from an auto-correlation length, $r_{0,\mathrm{QQ}}$, of $\approx8\,\cMpch$ at $z\approx2-3$ \citep[][]{white2012,eftekharzadeh2015}{}{} up to $\approx24\,\cMpch$ at $z\approx4$ \citep[][]{shen2007}{}{}. This rapid 
evolution in quasar clustering implies that quasars live in more massive halos as redshift increases, with a duty cycle that becomes larger as the number of host halos drops rapidly according to the exponential decline of the halo mass function \citep[][]{press_schechter1974}{}{}. At $z\approx4$, the situation seems to be particularly extreme, with host masses of $\gtrsim10^{13}\,\msun$ and a quasar lifetime approaching the Hubble time ($t_\mathrm{Q}\approx10^8-10^9\,\mathrm{yr}$) \citep[][]{shen2007, pizzati2023}{}{}. As highlighted by several studies \citep[][]{white_2008, wyithe_loeb2009, shankar_2010}{}{}, these values imply a steep and tight relation between the luminosity of quasars and the mass of the host halos, with SMBHs being either over-massive compared to their host halos/galaxies or having a large Eddington ratio 
\citep[][]{pizzati2023}{}{}. While these trends need to be backed up by the higher signal-to-noise measurements that will be allowed by future optical large-sky surveys, they paint a very interesting picture and call for studies of quasar clustering at even higher redshifts.

Measurements of the quasar auto-correlation function at $z\gtrsim5$, however, are extremely challenging due to the rapid decline of the quasar abundance at high redshift \citep[e.g.,][]{schindler2023}{}{}. One alternative pathway to determine the clustering of quasars is to cross-correlate them with some other tracer, e.g., coeval galaxies. The idea behind these measurements is that, if we assume that both quasars and galaxies trace the same underlying dark matter density distribution, but with different bias factors, the cross-correlation function between these two classes of objects is entirely determined by their respective auto-correlation functions. 
Given that the clustering of high-$z$ galaxies can be determined more easily due to their larger abundance, one can then measure the cross-correlation between quasars and galaxies (or, equivalently, study the over-densities of galaxies around quasars) to infer how strongly quasars are clustered in the high-$z$ Universe. 

Studies of the quasar-galaxy cross-correlation function are numerous at $z\approx0-5$, with results that overall confirm an increase in the clustering strength with redshift \citep[e.g.,][]{adelberger_steidel2005,shen2013,ikeda2015,garcia-vergara2017,he2018,garcia-vergara2019}{}{}. Nonetheless, two decades of ground- and space-based searches 
for galaxy over-densities around $z\gtrsim 6$ quasars have yielded mixed results, and contradictory claims have been made about the density (and clustering strength) of the primordial environments where these quasars live
\citep[e.g.][]{Stiavelli05,
Willott05,
Zheng06,
Kim09,
Morselli14,
Simpson14,Chiara17oden,Mignoli20}. In summary, even though the first studies on quasar clustering date back to more than two decades ago, extending these studies into the first billion years of cosmic history -- where the link between quasar clustering and SMBH growth is even more relevant -- has been extremely challenging. 

Recently, however, ground-breaking progress has been made following both of the two independent pathways mentioned above. Exploiting the high sensitivity of the Subaru High-$z$ Exploration of Low-Luminosity Quasars (SHELLQs) survey, \citet{arita2023} have compiled a sample of $\approx100$ faint quasars at $z\approx6$ and measured for the first time the large-scale quasar auto-correlation function at those redshifts. Despite the large uncertainties due to the limited size of their sample, the authors measured an auto-correlation length of $r_{0,\mathrm{QQ}}=24\pm11\,\cMpch$, in line with the trend observed at $z\approx4$. 

The launch of JWST, on the other hand, has opened up the possibility of obtaining large statistical samples of spectroscopically confirmed high-redshift galaxies, thus promising to revolutionize the search for over-densities around $z\approx6$ quasars. Indeed, several independent studies \citep[][]{kashino2022, aspire_wang2023}{}{} have already used NIRCam Wide Field Slitless Spectroscopic (WFSS) observations of $z\approx6$ quasar fields to show that these quasars reside in cMpc-scale over-densities traced by \OIII-emitting galaxies (\OIII {\it emitters}). Leveraging these unprecedented capabilities of JWST in studying the clustering and large-scale environment of high-redshift quasars, \citet[][hereafter \citeE24]{eilers2024} used observations from the EIGER survey \citep[][]{kashino2022,matthee23_eiger}{}{} to compile a catalog of \OIII emitters in the environments of four bright $z\approx6$ quasars, and measured for the first time the quasar-galaxy cross-correlation function at the same redshift. By also measuring the galaxy auto-correlation function, the authors concluded that high-$z$ quasars live on average in $\approx10^{12.3}\,\msun$ halos, although with a substantial quasar-to-quasar variance in terms of environments. This finding implies that $z\approx6$ quasars typically reside in moderately strong over-densities but not necessarily in the rarest and most massive environments that are present in the early Universe.

These measurements of the $z\approx6$ quasar auto-/cross-correlation functions offer a unique opportunity to study SMBHs and their properties at high-$z$. In \citet{pizzati2023} (hereafter, \citeP24), we showed that quasar clustering measurements can be combined with quasar demographic properties (expressed by the quasar luminosity function, QLF) to infer fundamental quantities such as the quasar luminosity-halo mass relation, the mass function of halos that host active quasars (the quasar-host mass function, QHMF), the quasar duty cycle and the quasar lifetime. \citeP24 makes use of a novel method that combines the outputs of dark-matter-only (DMO) cosmological simulations (specifically, the halo mass function and the cross-correlation function of halos with different masses) with an empirical quasar population model founded on a conditional luminosity function (CLF) framework \citep[e.g.,][]{yang2003}{}{}. The authors applied this model to measurements of the quasar auto-correlation and quasar luminosity functions at $z\approx2-4$, tracing the rapid change in SMBHs properties taking place at those redshifts.

In this work, we aim to extend the \citeP24 model to interpret the new measurements of the quasar-galaxy cross-correlation function and the auto-correlation functions of quasars and galaxies at $z\approx6$. These clustering measurements encompass a wide range of scales ($10^{-1}\lesssim r/\cMpc \lesssim 10^{3}$) and quasar luminosities ($10^{45.5}\lesssim L/\ergs\lesssim10^{48}$). Even more relevantly, modeling $z\approx6$ galaxies and quasars simultaneously to compute their cross-correlation statistics means that we must describe objects whose abundances span more than seven orders of magnitude 
\citep[][]{schindler2023, matthee23_eiger}{}{}. To overcome these obstacles, we extended the FLAMINGO suite \citep[][]{flamingo, FlamingoII}{}{} with a new $2.8\,\cGpc$ dark-matter-only simulation evolving more than a trillion particles and reaching the same resolution as the previous FLAMINGO DMO high-resolution runs \citep[][]{flamingo}{}{} but in a much larger volume. By employing this new, state-of-the-art, N-body simulation, named FLAMINGO-10k, we have the capability of modeling the clustering and demographic properties of quasars and galaxies simultaneously, providing a simple but powerful framework to interpret the large-scale environments of quasars and the properties of SMBHs in the first billion years of cosmic history.

The paper is structured as follows. In Sec. \ref{sec:methods}, we summarize the main features of the \citeP24 model and describe the improvements performed in this work. Sec. \ref{sec:population_model} lays down the general theoretical framework while the new FLAMINGO-10k simulation is described in Sec. \ref{sec:simulations}. Sec. \ref{sec:model-data} describes the comparison of our model with observational data, and Sec. \ref{sec:results} presents the main results of our analysis. These results are discussed and interpreted in the framework of current SMBH formation and evolution theories in Sec. \ref{sec:discussion}. Conclusions are provided in Sec. \ref{sec:conclusions}.

\section{Methods} \label{sec:methods}

The \citeP24 model takes two fundamental ingredients from cosmological simulations, i.e. the halo mass function and the cross-correlation functions of halos with different masses, and combines these with a quasar conditional luminosity function (which stochastically assigns quasars to halos) to reproduce observations of the quasar luminosity function and the quasar auto-correlation function, together with other relevant quantities such as the mass function of quasar-hosting halos and the quasar duty cycle (see their Fig. 1). 

Here, we plan to adapt this framework to include the presence of galaxies in the model, with the aim of reproducing their clustering and demographic properties in conjunction with the ones of quasars. We introduce the quasar-galaxy population modeling in Sec. \ref{sec:population_model} and Appendix \ref{sec:details_modeling}, and present the FLAMINGO-10k simulation on which the model is founded in Sec. \ref{sec:simulations}.

\subsection{Quasar and galaxy population models} \label{sec:population_model}

The primary goal of our model is to reproduce observations of the luminosity function and the clustering for both galaxies and quasars. 
In Appendix \ref{sec:details_modeling}, we outline a general framework that allows us to use a conditional luminosity function (CLF) to stochastically connect dark matter halos to any population of objects that are tracers of the underlying halo distribution and emit radiation with some luminosity, $L$. As discussed in the Appendix, both quasars and galaxies are suitable tracers to which this framework can be applied.
We do so simultaneously: we define a conditional luminosity function for quasars, CLF$_{\rm QSO}(L|M)$, and one for galaxies, CLF$_{\rm Gal}(L|M)$ -- with $L$ being the luminosity of quasars/galaxies and $M$ the mass of the host halos.

It is important to note that our definition of quasars and galaxies is entirely empirical, and it is solely based on our objective to reproduce a specific set of observations concerning these sources (see Introduction and Sec. \ref{sec:observations}). For this reason, our quasar population model is intended to describe only UV-bright, type-I quasars \citep[e.g.,][]{padovani2017}{}{}. As for galaxies, our objective is to match JWST observations of \OIII emitters (\citeE24), and thus -- when not explicitly stated otherwise -- we will use the words ``galaxies'' to describe only the ones that are bright in \OIII. Nonetheless, we stress the fact that the framework presented here is general and can be extended to different sub-populations of quasars/galaxies.

Another important note concerns the luminosity, $L$, of quasars and galaxies, which can also be set to any arbitrary choice (e.g., the bolometric luminosity or the luminosity of a specific line/band). 
As also done in \citeP24, we choose to work with bolometric luminosities when modeling quasars. Therefore, the quasar conditional luminosity function, CLF$_{\rm QSO}(L|M)$, will link the mass of host halos to the bolometric luminosities of quasars (i.e., $L\equiv L_\mathrm{bol}$). For galaxies, we use the luminosity of the \hbox{[O~$\scriptstyle\rm III $]}$_{5008}$ line, $L_\mathrm{OIII}$ instead, as this is the quantity that determines the detectability of the galaxies in the  (slitless) JWST surveys.
Therefore, CLF$_{\rm Gal}(L|M)$ relates halos to \OIII luminosities (i.e., $L\equiv L_\mathrm{OIII}$). In the following, we will always use the symbol $L$, but add the caveat that the specific value of this symbol is different depending on whether we refer to quasars or galaxies.

We assume the same functional form for the two conditional luminosity functions, CLF$_{\rm QSO}$ and CLF$_{\rm Gal}$. Following \citeP24, we write\footnote{As also discussed in \citeP24, the factor $f_{\rm on}$ accounts for the fact that not all quasars/galaxies may be luminous at any given time. In other words, we are implicitly assuming that a fraction of sources are inactive or simply too dim to be revealed by any observations and therefore we do not include their contribution in the CLF.} 
\begin{equation}
        {\rm CLF}_i(L| M)\, \d L= \,\frac{f_{{\rm on}}^{(i)}}{\sqrt{2\pi}\sigma^{(i)}}\,e^{-\frac{\left(\log_{10} L - \log_{10} L_{\mathrm{c}}^{(i)}(M)\right)^2}{2\sigma^{(i)2}}} \d \log_{10} L ,
        \label{eq:clf_log_normal}
\end{equation}
where $i$ stands either for ``QSO'' or ``Gal''. 
The characteristic luminosity, $L_\mathrm{c}^{(i)}$, has a power-law dependence on halo mass:
\begin{equation}
    L_{\mathrm{c}}^{(i)}(M) = L_{{\rm ref}}^{(i)} \,\left(\frac{M}{M_{\rm ref}}\right)^{\gamma^{(i)}}. \label{eq:power_law_Lc}
\end{equation}
with $M_{\rm ref}$ being a reference mass that is associated with the reference luminosity $L_{\rm ref}$; we fix it to $\log_{10} M_{\rm ref}/\msun = 12.5$. The free
parameters of the model, which we will infer directly from observations in Sec. \ref{sec:observations}-\ref{sec:results}, are $\sigma^\mathrm{(QSO,Gal)}$, $L_\mathrm{ref}^\mathrm{(QSO,Gal)}$, $\gamma^\mathrm{(QSO,Gal)}$, and $f_\mathrm{on}^\mathrm{(QSO,Gal)}$.
Note that, as in \citeP24, we assume that these parameters do not depend on other variables such as halo mass or quasar luminosity.

Using the general framework outlined in Appendix \ref{sec:details_modeling} (see also \citeP24), we can combine each conditional luminosity function, CLF$_{\rm QSO}$ and CLF$_{\rm Gal}$, with the halo mass function, $n_{\rm HMF}$, to obtain fundamental quantities describing quasars and galaxies, such as their luminosity functions ($n_{\rm QLF}$ and $n_{\rm GLF}$), host mass functions ($n_{\rm QHMF}$ and $n_{\rm GHMF}$), and duty cycles ($\varepsilon_{\rm QDC}$ and $\varepsilon_{\rm GDC}$). 

The quasar luminosity function (QLF) and the galaxy luminosity function (GLF) are observable quantities, and hence the predictions from our model for these functions can be directly compared with data. As for the quasar-host mass function (QHMF) and the galaxy-host mass function (GHMF), they determine the clustering properties of quasars and galaxies, respectively. 

In particular, we follow here the approach described in \citeP24 (see their Section 1 and Appendix A) to write the clustering properties of a population of objects given its host halo mass distribution. This approach assumes that the cross-correlation functions of dark matter halos with different masses are known. We describe in Sec. \ref{sec:simulations} and Appendix \ref{sec:details_fitting} how to extract these cross-correlation terms from a cosmological simulation. Here, we assume that, after creating bins in halo mass, we can write the cross-correlation between two mass bins as $\xi_h(M_j, M_k; r)$, with $M_{j,k}$ being the bin centers. 

The point made in \citeP24 is that all the correlation functions concerning quasars and galaxies are simply weighted averages of these cross-correlation terms, with the weights ($Q_j,G_j$) determined by the specific host mass distribution we are considering ($n_\mathrm{QHMF}$ for quasars and $n_\mathrm{GHMF}$ for galaxies).
In particular, we can define the weights $Q_j$ to be:
\begin{equation}
     Q_j = \frac{n_\mathrm{QHMF}( M_j|L>L_{\rm thr})\,\Delta M }{\int_0^{M_{\rm max}} n_\mathrm{QHMF}( M|L>L_{\rm thr})\,\d M}, \label{eq:weight_Qj}
\end{equation}
with $\Delta M$ being the width of the mass bins. The identical weighting for galaxies, $G_j$, reads:
\begin{equation}
     G_j = \frac{n_\mathrm{GHMF}( M_j|L>L_{\rm thr})\,\Delta M }{\int_0^{M_{\rm max}} n_\mathrm{GHMF}( M|L>L_{\rm thr})\,\d M}. \label{eq:weight_Gj}
\end{equation}
With these definitions, we can write all correlation functions in the general form (with A and B representing two different populations of halo tracers):
    \begin{equation}
        \xi_\mathrm{AB}(r) 
        = \sum_{j,k} A_j B_k \xi_h(M_j, M_k; r). \label{eq:corr_func}
    \end{equation}
This expression implies that the quasar auto-correlation function, $\xi_\mathrm{QQ}(r)$, can simply be written as:
\begin{equation}
        \xi_\mathrm{QQ}(r) 
        = \sum_{j,k} Q_j Q_k \xi_h(M_j, M_k; r), \label{eq:quasar_corr_func}
    \end{equation}
with the weights set by eq. \ref{eq:weight_Qj}.
In the same way, the galaxy auto-correlation function, $\xi_\mathrm{GG}(r)$, reads:
\begin{equation}
        \xi_\mathrm{GG}(r) 
        = \sum_{j,k} G_j G_k \xi_h(M_j, M_k; r), \label{eq:galaxy_corr_func}
    \end{equation}
Finally, the cross-correlation function between quasars and galaxies, $\xi_\mathrm{QG}(r)$, is retained by weighting over the QHMF and the GHMF simultaneously:
\begin{equation}
        \xi_\mathrm{QG}(r) 
        = \sum_{j,k} Q_j G_k \xi_h(M_j, M_k; r). \label{eq:cross_corr_func}
    \end{equation}
    
As a final step, all of these correlation functions can be integrated along the line of sight direction to average out the contribution of redshift space distortions. In this way, we compute quantities that can be directly matched with data, such as the projected correlation function, $w_p(r_p)$, or the volume-averaged correlation function, $\chi_V(r_p)$. 
The former follows from a simple integration along the line of sight direction, $\pi$, with a limit $\pi_\mathrm{max}$ that is chosen according to observations:
    \begin{equation}
        w_p(r_p) = 2\int_{0}^{\pi_{\rm max} } \xi(r_p, \pi)\,\d \pi ,
        \label{eq:projected_corrfunc}
    \end{equation}
while the latter implies that we choose a radial binning in the perpendicular direction, $r_p$, and a maximum distance in the parallel direction, $\pi_{\rm max}$, and perform a spatial average of the correlation function on every cylindrical bin. If we define $r_{p,\mathrm{min}}$ and $r_{p,\mathrm{max}}$ as the lower and upper limits of the radial bins, respectively, $\chi_V(r_p)$ can be simply expressed as:
   \begin{equation}
        \chi_V(r_p) = 
    \frac{2}{V}\int_{r_{p,\mathrm{min}}}^{r_{p,\mathrm{max}}}\int_{0}^{\pi_{\rm max}}\xi(r_p, \pi)\,2\pi r_p\,{\rm d}r_p\,{\rm d}\pi. 
        \label{eq:projected_corrfunc_vol}
    \end{equation}

\subsection{Simulation setup} \label{sec:simulations}

As described in \citeP24 (see their Figure 1), we use dark-matter-only (DMO) cosmological simulations to extract two fundamental quantities that are at the core of our model: the \textit{halo mass function}, $n_\mathrm{HMF}$, and the \textit{cross-correlation functions} of halos with masses $M_j$ and $M_k$, $\xi_h(M_j,M_k;r)$. 

\citeP24 used multiple simulations with different box sizes and resolutions to extend the range of masses that can be reliably modeled in their framework. The argument in support of this approach was that every different simulation can describe the demographic and clustering properties of halos in a different range of masses, and putting together these properties allows for an exploration of a larger set of quasar-host mass distributions. This approach was particularly suited for getting an estimate of the quasar auto-correlation function, as this quantity primarily depends on the auto-correlation function of the halos whose mass 
is the maximum of the QHMF. For this reason, resolving very low and very high mass halos in the same simulation was not necessary, and the terms of the cross-correlation functions $\xi_h(M_i,M_j;r)$ with, e.g., $M_i\gg M_j$ were just extrapolated by appropriate analytic functions (see \citeP24 for more details).

The problem we are facing here, however, is intrinsically different, as we need to model the cross-correlation function between quasars -- which are very rare and are expected to live in massive halos -- and galaxies -- which are much more abundant and hence are hosted by much more common systems. This implies that the cross-correlation functions between very massive and less massive halos are at the core of our model, and hence they need to be faithfully represented in our numerical setup. For this reason, we use here a single simulation with a larger number of particles, intending to represent in the same box halos whose range of masses is broad enough to account for the presence of quasars and galaxies simultaneously. In the following, we give more details about the properties of this simulation, and we then proceed to describe how we extract from the simulated box the halo properties that our population models require. 

\subsubsection{Extending the suite of FLAMINGO runs: FLAMINGO-10k}

FLAMINGO \citep[][]{flamingo, FlamingoII} is a suite of state-of-the-art, large-scale structure cosmological simulations combining hydrodynamical and dark-matter-only (DMO) runs in large volumes ($\geq1~\rm{Gpc}$). The simulations were performed using the coupled Particle-Mesh \& Fast-Multipole-Method code \code{SWIFT} \citep[][]{swift}{}{}. The fiducial runs adopt the ``3x2pt + all'' cosmology from \citet{abbott_des2022} ($\Omega_\mathrm{m} = 0.306$, $\Omega_\mathrm{b} = 0.0486$, $\sigma_8 = 0.807$, $\mathrm{H}_0 = 68.1\,\kms\,{\rm Mpc}^{-1}$, $n_\mathrm{s} = 0.967$), with a summed neutrino mass of $0.06\,\mathrm{eV}$. 
Initial conditions (ICs) are set using multi-fluid third-order Lagrangian perturbation theory (3LPT) implemented in \textsc{MonofonIC} \citep[][]{hahn2020monofonic,michaux2021monofonic}{}{}. Partially fixed ICs are used to limit the impact of cosmic variance \citep[][]{angulo_pontzen2016}{}{} by setting the amplitudes of modes with wavelengths larger than $1/32$ of the simulation volume side-length to the mean. The most demanding simulation in the suite (the \texttt{L2p8\_m9} run of \citealt{flamingo}) encompassed a volume of side-length $2.8~\rm{cGpc}$ with particles of mass $6.72\times10^9\,\msun$.

Whilst the volume of this flagship run is sufficient for the present study, the resolution is not high enough to reliably characterize the halo mass and clustering of the [OIII] emitters we seek to study. We thus ran an additional simulation, FLAMINGO-10k, which we add to the FLAMINGO suite. FLAMINGO-10k was run on 65\ 536 compute cores, using the same setup (software, cosmology, ...) as the previous DMO FLAMINGO simulations, but with 8x higher resolution than the \texttt{L2p8\_m9} run and a higher starting redshift ($z=63$). The box size of this new simulation is chosen according to the flagship FLAMINGO run, $L=2.8\,\cGpc$, while the resolution of the simulation reaches the one of the $1\,\cGpc$ FLAMINGO DMO high-resolution run ($m_\mathrm{CDM}=8.40\times10^{8}\,\msun$). The simulation makes use of $10080^3$ cold dark matter (CDM) particles and $5600^3$ neutrino particles, resulting in a total number of particles close to $1.2\times10^{12}$. As detailed in Sec.~\ref{sec:results}, this large number of particles will let us model halos whose masses span more than two orders of magnitude at $z\approx6$ throughout the $(2.8\,\cGpc)^3$ volume. The particles and halo catalogs were stored at 145 redshifts between $z=30$ and $z=0$ with $31$ outputs at $z>6$, allowing for the precise tracing of the growth of structures at early times.

\subsubsection{Obtaining the sub-halo catalogue with HBT+}\label{sec:hbt}

The first step that we take once we have the final simulated volume is to build a halo catalogue containing the positions and masses of all (sub-)halos in the simulation. In \citeP24, we included only central halos in the catalogue and discarded the contribution of satellite haloes completely. This was done because our main focus was the auto-correlation function of quasars at large scales ($r\gtrsim5\,\cMpc$). Here, instead, we aim to reproduce correlation functions down to $r\approx0.1\,\cMpc$ (i.e., well within the virial radii of massive halos), and hence the contribution of all sub-haloes must be carefully considered. We note that in our framework (Sec. \ref{sec:population_model}) we do not make any explicit distinctions between central sub-halos and satellites. For this reason, we build a halo catalogue that includes all kinds of sub-halos, and we use the general term ``halo'' to refer to any kind of sub-halos, irrespective of whether they are central or satellite. In general, whenever we refer to quasar/galaxy hosts in the context of our model (e.g., in the QHMF and GHMF), we always implicitly assume that we are talking about \textit{sub-halos}, and not about the larger groups identified by a friends-of-friends algorithms.

We select a single snapshot from FLAMINGO-10k at $z=6.14$, which represents the closest match in terms of redshift to the observations we aim to reproduce in this work (Sec. \ref{sec:observations}).
We use this snapshot together with all the other ones at higher-$z$ to build a halo catalogue using the upgraded Hierachical Bound-Tracing (HBT+) code \citep[][]{hbt, hbt_plus}{}{}. HBT+ identifies sub-haloes as they form and tracks their evolution as they merge. By consistently following sub-haloes across cosmic times HBT+ represents a robust solution to the problem of identifying small-scale bound
structures 
in DMO simulations. This is the ideal choice for the problem we are facing here, as we aim to represent the spatial distribution of quasars and galaxies down to very small spatial scales.

We use the bound mass definition for (sub-)halo masses. In other words, we compute the mass of each (sub-)halo by summing up the mass of all its bound particles. Since tidal stripping decreases the mass of satellite halos by a significant amount, we use here the peak halo mass, $M_\mathrm{peak}$, which is defined as the largest bound mass that a (sub-)halo has had across cosmic history.  
In practice, HBT+ saves this mass for each snapshot, and so we can simply use the peak bound masses that are given in the output by the code for our population model (i.e., $M\equiv M_\mathrm{peak}$). We then complete the catalogue by adding the position of each (sub-)halo, which we define by looking at its centre of potential.

\subsubsection{A simulation-based analytical description of halo properties} \label{sec:dmo_fit}

Once we have obtained a catalogue with the positions and masses of halos in the simulation at a given redshift, we can easily compute the halo mass function and the (cross-)correlation functions of halos with different masses. However, as also done in \citeP24, we aim to describe these quantities with analytical functions, which we fit to the outputs of the simulation. This approach allows us to obtain a very general description of halo properties, independent of the specific mass bins employed. More importantly, in \citeP24 we have shown that using these fitting functions we can smoothly extrapolate the behavior of the cross-correlation functions even to the combinations of mass bins for which there are very few halos available in the simulation, and hence for which the correlation functions measured numerically are extremely noisy and uncertain. This simple step improves the quality of our parameter inference (Sec. \ref{sec:model-data}) and lets us recover well-behaved posterior distributions for a wide range of model parameters. 

Fitting the halo mass function is straightforward. As in \citeP24, we consider the same functional form used by \citet{tinker2008} (see also \citealt{jenkins2001,white2002, warren2006}) for the fit, and consider all halos above the minimum mass $\log M_\mathrm{min}/\msun = 10.5$, corresponding to halos with more than $\approx40$ particles.

As for the cross-correlation function of halos with mass $M_j$ and $M_k$, $\xi_h(M_j, M_k; r)$, we first compute each correlation function numerically by creating a grid in mass and distance made by 8 uniformly spaced bins in $\log_{10} M$, 
with a minimum halo mass of $\log_{10} M_{\rm min}/\msun=10.5$ and a maximum of $\log_{10} M_{\rm max}/\msun=12.5$, and 18 (logarithmically-spaced) bins in the radial direction with a minimum radial distance of $\log_{10} r_{\rm min}/\cMpc = -1$ and a maximum of $\log_{10} r_{\rm max}/\cMpc = 2.2$.
We then use the package \code{corrfunc} \citep[][]{corrfunc} to compute the number of halo pairs in the simulated catalogues for every combination of masses and distance. We use a simple estimator to obtain the halo cross-correlation functions:
\begin{equation}
    \xi_h(M_j, M_k; r) = \xi_{j,k}(r) = \frac{D_jD_k(r)}{R_jR_k(r)}-1, \label{eq:pair_counts}
\end{equation}
where $D_jD_k$ stands for the number of pairs of halos in the mass bin $j$ with halos in the mass bin $k$, whereas $R_jR_k$ refers to the number of pairs when comparing to a random distribution of the same halos. For a periodic box of volume $V$, $R_jR_k$ can be simply expressed analytically as:
\begin{equation}
    R_jR_k = \frac{4\pi}{3V}\left(r_\mathrm{max}^3-r_\mathrm{min}^3\right) \,N_jN_k ,
\end{equation}
with $N_j$ and $N_k$ being the number of halos in the mass bins $j$ and $k$, respectively, and $r_\mathrm{min,max}$ the limits of the radial bin considered.

We fit $\xi_h(M_j, M_k; r)$ with the same setup as described in \citeP24. In short, we divide all the cross-correlation terms, $\xi_h(M_j, M_k; r)$, by a reference correlation function, $\xi_{\rm ref}(r)$, which we set equal to the auto-correlation function of the first mass bin. Then, we fit the resulting functions with a 3-d polynomial to capture the residual dependencies on the two masses and the distance. The fit is performed by converting masses to peak heights, $\nu(M) = \delta_c/\sigma(M,z)$ -- with $\delta_c\approx 1.69$ and $\sigma^2(M,z)$ being the variance of the smoothed linear density field (see also Sec. \ref{sec:discussion_redshift}). We adopt this approach in order to minimize any dependences of the cross-correlation functions on cosmology and redshift. Errors on the cross-correlation terms are chosen by assuming Poissonian uncertainties on the halo pair counts. Finally, we note that, before fitting, we weigh 
every uniform mass bin with the halo mass function, so that the effective mass $M_k$ corresponding to the bin $k$ is not the bin center, but the median value of the halo mass function in that specific bin. 

After performing the fit, we introduce here a further step that aims to achieve a better description of the cross-correlation functions at large scales, $r\gtrsim20-40\,\cMpc$. As noted in \citeP24, the values of the correlation functions extracted from simulations tend to be unreliable at large scales for two reasons. First, the finite size of the box reduces the number of very large-scale pairs that are available. Secondly, at $r\gtrsim100\,\cMpc$ the behavior of correlation functions becomes non-trivial due to the presence of the baryon acoustic oscillations (BAO) peak, which is hard to capture with the coarse binning employed here. At large scales, however, density perturbations are linear and they can faithfully be described by the linear halo bias framework  \citep[][]{bardeen1986,cole_kaiser1989,jing98,cooray_sheth2001}. For this reason, we follow \citet[][]{darkquest}{}{} and smoothly interpolate between our fit to simulations at small-to-medium scales and the predictions from linear theory at large scales. In practice, we introduce a damping function $D(r)$, and write the correlation functions $\xi_h(M_j, M_k; r)$ as:
\begin{equation}
    \xi_h(M_j, M_k; r) =  D(r) \xi_{h,\mathrm{fit}}(M_j, M_k; r) + (1-D(r))  \xi_{h, \mathrm{lin}}(M_j, M_k; r),
\end{equation}
where $\xi_{h,\mathrm{fit}}(M_j, M_k; r)$ is the fit performed to simulations described above, while $\xi_{h, \mathrm{lin}}(M_j, M_k; r)$ is the prediction coming from the linear halo bias framework (based on linear theory, see, e.g., \citealt{halomod}): 
\begin{equation}
    \xi_{h, \mathrm{lin}}(M_j, M_k; r) = b(M_j) b(M_k) \xi_\mathrm{mm}(r). \label{eq:corr_linear}
\end{equation}
We use the package \code{colossus} \citep[][]{colossus_diemer2018}{}{} to compute the matter auto-correlation function, $\xi_\mathrm{mm}(r)$, and the linear bias factors, $b(M_{j,k})$, based on the \citet[][]{tinker2010}{}{} relation.
As for the damping function, we choose the following functional form:
\begin{equation}
    D(r) = e^{-\left(\frac{r}{r_\mathrm{lin}}\right)^\alpha},
\end{equation}
with the parameters set to $\alpha=5$ and $r_\mathrm{lin}=20\,\cMpc$.

In summary, we adopt here an extension of the \citeP24 fitting framework that uses DMO simulations to provide an analytical description of the demographic and clustering properties of halos, expressed by the halo mass function and the halo cross-correlation functions. Thanks to the use of fitting functions, we can extrapolate the behavior of these quantities for a very large range of masses (from $\log M/\msun\approx10.5$ to $\log M/\msun\approx13-13.5$), and, by smoothly interpolating between DMO simulations at small scales and linear theory at large scales, our correlation functions can capture more than four orders of magnitude in scale (from $r\approx0.1\,\cMpc$ out to $r\approx1\,\cGpc$). As shown in the following Section, these properties are essential to reproduce the large diversity of data concerning galaxies and quasars that are the focus of the present work. 

In Appendix \ref{sec:details_fitting}, we show the results for the fit of the cross-correlation function terms and elaborate on the validity of this approach in the context of our analysis. Further discussion on the general methodology employed here can be found in \citeP24.

\section{Data-Model comparison} \label{sec:model-data}

\begin{table*}
\centering
\setlength{\extrarowheight}{3pt}
\caption{Summary of all the data we compare our model with, together with a quantitative measurement ($\chi^2$ statistics) of the quality of the fit. The $\chi^2$ is computed by considering the best-fit parameters coming from the joint fit (see main text), and $n$ stands for the number of data points that we fit for each quantity. A discussion on the quality of the fit can be found in Sec. \ref{sec:parameter_inference}.}
\begin{threeparttable}
\begin{tabular}{c | c c c c c | c }
\toprule
 Name & Quantity & Survey Name & Redshift Range & Figure & Reference &  $\chi^2/n$\\ 
\midrule
Quasar luminosity function &
$n_\mathrm{QLF}(L)$ &
PS1, SHELLQs &
$5.7-6.2$ &
Fig. \ref{fig:overview} (top) &
\citet[][]{schindler2023}{}{}     &   
$8.3/10$ \\
Galaxy luminosity function\tnote{a} &
$n_\mathrm{GLF}(L)$ &
EIGER &
$5.3-6.9$ &
Fig. \ref{fig:overview} (top) &
\citet[][]{matthee23_eiger}{}{}     &   
$8.7/7$ \\
\midrule
Quasar-Galaxy cross-correlation function &
$\chi_{V,\mathrm{QG}}(r_p)$ &
EIGER &
$5.9-6.4$ &
Fig. \ref{fig:overview} (bottom) &
\citeE24     &   
$7.4/8$ \\
Galaxy auto-correlation function &
$\chi_{V,\mathrm{GG}}(r_p)$ &
EIGER &
$5.3-6.9$ &
Fig. \ref{fig:overview} (bottom) &
\citeE24     &   
$15.1/8$ \\
Quasar auto-correlation function &
$w_{p,\mathrm{QQ}}(r_p)/r_p$ &
SHELLQs &
$5.8-6.6$ &
Fig. \ref{fig:autos_arita} &
\citet[][]{arita2023}{}{}     &   
$6.1/5\tnote{b}$ \\
\bottomrule
\end{tabular}
\begin{tablenotes}
    \item [a] We exclude the innermost bin because it is very uncertain due to low completeness.
    \item [b] This dataset is excluded from the joint fit, and analysed separately in Appendix \ref{sec:autos_arita}. The $\chi^2$ reported here is the value obtained using the best-fit parameters coming from the joint fit of all the other datasets.
\end{tablenotes}
\label{tab:obs_data}
\end{threeparttable}
\end{table*}

Adopting the methodology described in the previous Section, we can obtain all the ingredients needed to compare our model with observational data. The model depends on eight free parameters (see Sec. \ref{sec:population_model}), 
that we constrain by jointly fitting the luminosity and clustering measurements of both quasars and galaxies. We provide a brief description of the data considered in the analysis in Sec. \ref{sec:observations}, and proceed to the comparison with our model in Sec. \ref{sec:parameter_inference}.

\subsection{Overview of observational data}
\label{sec:observations}

The data we consider in this work concern the luminosity functions and auto-correlation functions of quasars and galaxies, and the cross-correlation function between these two different populations. 
In Table \ref{tab:obs_data}, we summarize all these data and point to their respective references. The $z\approx6$ quasar luminosity function (QLF) is taken from \citet[][]{schindler2023}{}{}, and it is compiled including 125 quasars with $-28 \lesssim M_{1450} \lesssim -25$ from the the  Pan-STARRS1 (PS1) quasar survey \citep[][]{banados2016}{}{}, as well as 48 fainter ($-25 \lesssim M_{1450} \lesssim -22$) quasars from the SHELLQs survey \citep[][]{kashikawa2015,matsuoka2018}{}{}. Note that, as detailed in Sec. \ref{sec:population_model} and in \citeP24, we convert absolute magnitudes to quasar bolometric luminosities using the relation from \citet{runnoe2012}\footnote{The bolometric correction for $\lambda=1450$ \r{A} 
is $\log_{10} L_{\rm iso}/\ergs  = 4.745 + 0.910 \log_{10} \lambda L_\lambda/\ergs$. $L_{\rm iso}$ refers to the bolometric luminosity computed under the assumption of isotropy, and it is related to the real bolometric luminosity $L$ through the relation $L = 0.75 \, L_{\rm iso}.$}. The galaxy luminosity function (GLF), based on JWST observations of \OIII emitters, was compiled by \citet[][]{matthee23_eiger}{}{} in the context of the EIGER survey. The luminosities of galaxies are already expressed in \OIII line fluxes, in accordance with our population model (Sec. \ref{sec:population_model}), and cover the range $42\lesssim \log_{10} L_\mathrm{OIII}/\ergs \lesssim 43.5$. We discard the faintest bin in the GLF because, as discussed in \citet[][]{matthee23_eiger}{}{}, its completeness is relatively low ($\approx40\%$), and hence the value of the abundance of galaxies in that bin is 
particularly uncertain. 

The quasar-galaxy cross-correlation function and the galaxy auto-correlation function are also measured by the EIGER survey in \citeE24. They both span a spatial range $0.1\lesssim r/\cMpc \lesssim 6$, sharing the same radial bins. Being obtained with the same methodology and in the same analysis, these two datasets are homogeneous, and it is natural to consider them jointly. The quasar auto-correlation function \citep[][]{arita2023}{}{}, on the other hand, comes from a very different dataset: it includes quasars with much fainter luminosities from the SHELLQs survey \citep[][]{matsuoka2018}{}{}, and it constrains their clustering only at very large scales ($r\gtrsim40\,\cMpc$; see \citealt{arita2023}).
Further discussion on this can be found in Sec. \ref{sec:parameter_inference} and in Appendix \ref{sec:autos_arita}.  

One of the key aspects to bear in mind when analysing data concerning correlation functions is that our model is quite sensitive to the value of the luminosity threshold, $L_{\rm thr}$, considered when measuring quasar/galaxy clustering (see eq. \ref{eq:qhmf}-\ref{eq:hod}). While properly modeling the effects of observational incompleteness in the context of our framework is beyond the scope of this work, it is important to set these threshold values carefully to ensure that we get unbiased results. Let us start with the \citeE24 observations. The EIGER survey targets only five 
very bright quasars and detects galaxies in their fields. This implies that the quasar population whose clustering is being probed by EIGER consists only of very bright ($M_{1450}\lesssim-27$) sources. For this reason, we set a value of the quasar luminosity threshold for modeling the quasar-galaxy cross-correlation function of $\log_{10} L_\mathrm{thr,QSO}/\ergs=47.1$, which is consistent with the luminosity of the faintest quasar probed by EIGER. However, we mention the caveat that setting a luminosity threshold would only be possible for a luminosity-limited sample.
In reality, the EIGER survey targets only a few selected quasar fields and is not constructed to reproduce the actual luminosity distribution of bright quasars. While this may introduce a minor bias in our results, we neglect this effect here and consider the EIGER sample to be representative of the $z\approx6$ bright ($L>L_\mathrm{thr,QSO}$) quasar population.

As for galaxies, the minimum \OIII luminosity that EIGER measurements consider is $\log_{10} L/\ergs\approx42$. However, the sample starts to be significantly incomplete already at higher luminosities. This represents an issue in our framework, 
as the luminosity-halo mass relations assumed in eq. \ref{eq:power_law_Lc} imply that clustering is luminosity-dependent. Including a large population of low-luminosity galaxies of which only a fraction was detected in the observations because of low completeness would then bias our results, since the luminosity distribution of the galaxies for which clustering was measured would not be the same as the
one resulting from our modeling by simply setting the luminosity limit to be the lowest luminosity considered.
We can alleviate this problem by setting an effective luminosity threshold that accounts for the fact that the sample is largely incomplete at lower luminosities.   
We choose the following effective threshold for galaxies: $\log_{10} L_\mathrm{thr,Gal}/\ergs\approx42.4$. This value corresponds to the luminosity at which the average completeness of the EIGER sample drops below $\approx75\%$ \citep[][]{matthee23_eiger}{}{}.
We employ an analogous argument to set the luminosity threshold for the quasar auto-correlation function measurements of \citet[][]{arita2023}{}{}. We find the magnitude at which the completeness of the SHELLQs survey drops below $75\%$, and convert this magnitude to a quasar bolometric luminosity obtaining $\log_{10} L_\mathrm{thr,QSO}/\ergs=45.3$\footnote{As detailed in Sec. \ref{sec:parameter_inference} and Appendix \ref{sec:autos_arita}, we find that the data for the quasar auto-correlation function are not able to constrain our model parameters. For this reason, in this specific case, the value for the luminosity threshold we choose here is irrelevant.}.

\begin{figure*}
	\centering
	\includegraphics[width=0.53\textwidth]{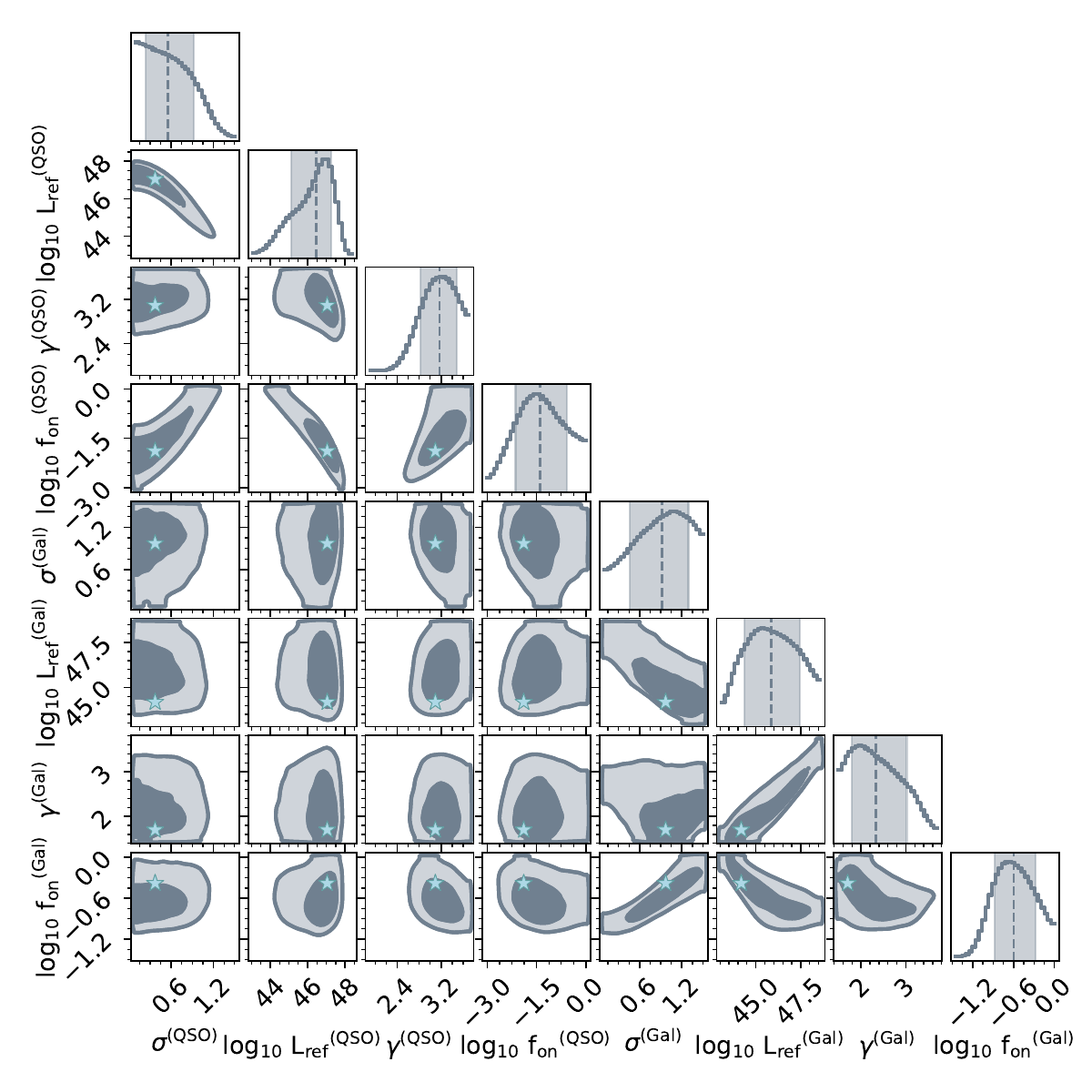}
 	\includegraphics[width=0.46\textwidth]{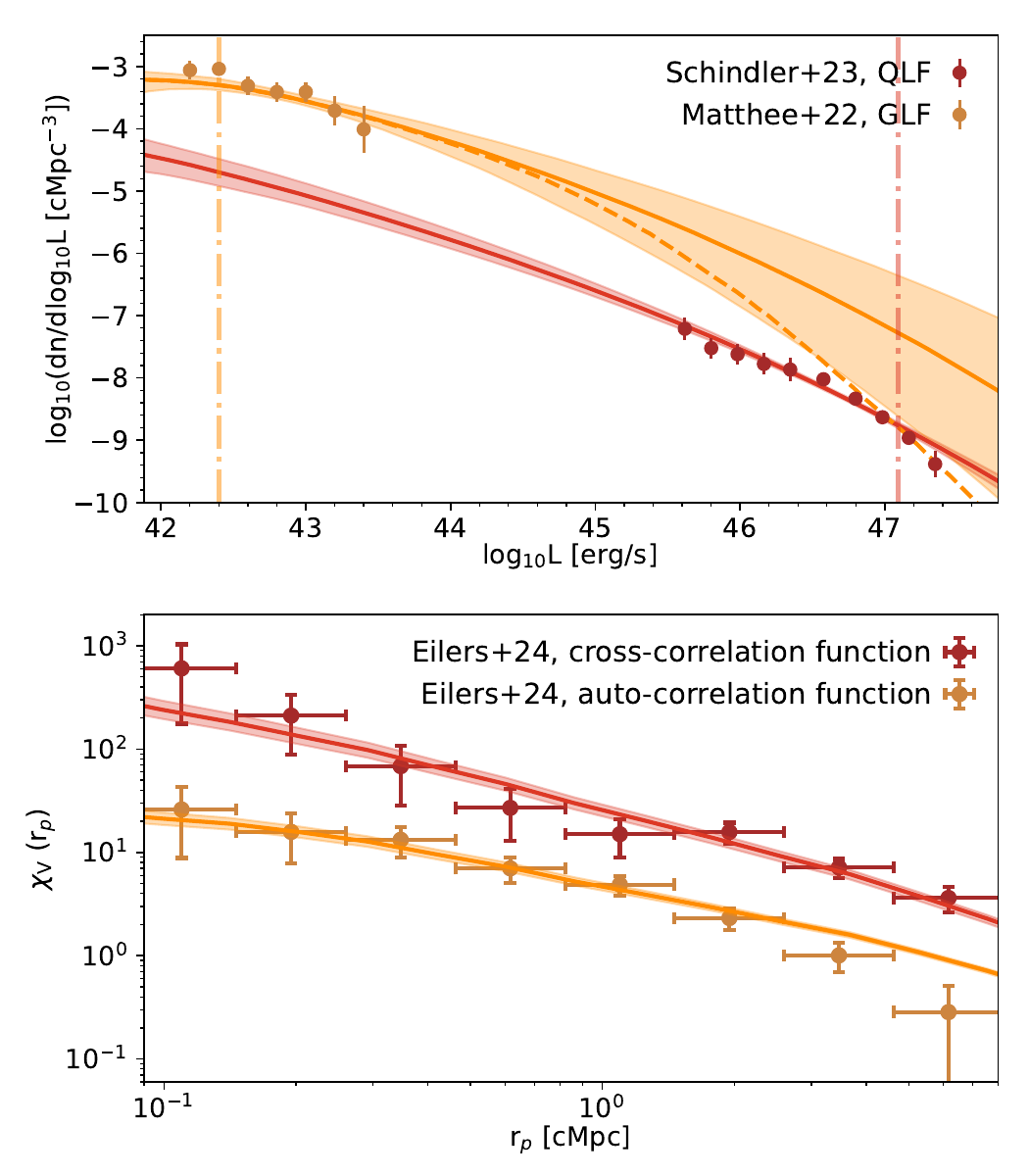}

	 \caption{\textit{Left}:
	Corner plots of the 8-d posterior distribution for the joint fit described in Sec. \ref{sec:parameter_inference}. 
	Contours in the 2-d histograms highlight the $1\sigma$ and $2\sigma$ regions, whereas the dashed lines in the 1-d histograms represent the median values of the parameters (with $1\sigma$ errors shown as shaded regions).
    The maximum-likelihood values are shown with star symbols in each corner plot. The units of the reference luminosity parameters $\log_{10} L_{\rm ref}^\mathrm{(QSO,Gal)}$ are $\ergs$.
	\textit{Right}: Comparison of the predicted luminosity (top) and correlation (bottom) functions with the observational data from Table \ref{tab:obs_data}. The galaxy luminosity function (GLF) and auto-correlation function are shown in orange, while the quasar luminosity function (QLF) and the quasar-galaxy cross-correlation function are shown in red. Median values (solid lines) and $1\sigma$ uncertainty regions (shaded areas) are obtained by randomly sampling the Markov chains for the posterior distribution $2000$ times. 
    The red and orange vertical dot-dashed lines in the upper right panel are the luminosity threshold for quasar and galaxies ($L_{\rm thr}$), respectively, that are used for modeling clustering measurements (see Sec. \ref{sec:model-data}). The dashed line in the same panel represents the median value for the GLF when assuming that the galaxy luminosity-halo mass relation flattens for large halo masses (see Sec. \ref{sec:results} and Figure \ref{fig:results_qg}).
  \label{fig:overview}
 	}
\end{figure*}

\subsection{Parameter inference}
\label{sec:parameter_inference}

\begin{table}
\centering
\setlength{\extrarowheight}{3pt}
\caption{Constraints (median values and 16th-84th percentiles) on the model parameters based on the corner plots shown in Figure \ref{fig:overview}. The eight parameters are divided between the ones describing the quasar CLF (``QSO'') and the ones for the galaxy CLF (``Gal'').}

\begin{tabular}{c | c c c c }
\toprule
 Quantity & $\sigma$ & $\log_{10} L_{\rm ref}$ [$\ergs$]    & $\gamma$  & $f_{\rm on}$ [\%]\\ 
\midrule
QSO &
$ 0.55_{-0.31}^{+0.37} $ &
$ 46.45_{-1.35}^{+0.79} $      &   
 $ 3.17_{-0.34}^{+0.32} $  &
 $ 3.9_{-3.2}^{+21} $ \\
\midrule
Gal. &
$ 0.92_{-0.46}^{+0.38} $  &
$ 45.86_{-1.49}^{+1.60} $      &   
$ 2.33_{-0.54}^{+0.69} $ &
$ 25_{-20}^{+31} $ \\
\bottomrule
\end{tabular}
\label{tab:mcmc_results}
\end{table}

We employ a Bayesian framework and write down the posterior distribution for the model parameters. As described in Sec. \ref{sec:population_model}, the model has eight free parameters, describing the conditional luminosity functions of quasars and galaxies simultaneously. We choose the same parametrization for $\mathrm{CLF}_\mathrm{QSO}$ and $\mathrm{CLF}_\mathrm{Gal}$. As a result, the same sets of parameters account for the two functions: these are the normalization and slope of the quasar/galaxy luminosity-halo mass relation ($L_{\rm ref}$ and $\gamma$, respectively), the logarithmic scatter around this relation ($\sigma$), and the fraction of quasars/galaxies that are active at any given moment ($f_{\rm on}$). The final set of parameters, $\Theta$, is then: $(\sigma^\mathrm{(QSO)}, L_\mathrm{ref}^\mathrm{(QSO)}, \gamma^\mathrm{(QSO)}, f_\mathrm{on}^\mathrm{(QSO)}, \sigma^\mathrm{(Gal)}, L_\mathrm{ref}^\mathrm{(Gal)}, \gamma^\mathrm{(Gal)}, f_\mathrm{on}^\mathrm{(Gal)})$.

As in \citeP24, we set flat uninformative
priors on $\sigma^\mathrm{(QSO,Gal)}$ and $\gamma^\mathrm{(QSO,Gal)}$, and on the logarithm of $L_\mathrm{ref}^\mathrm{(QSO,Gal)}$ and $f_\mathrm{on}^\mathrm{(QSO,Gal)}$. We choose to explore a wide parameter space, letting the parameters vary with the following bounds:
$\sigma^\mathrm{(QSO,Gal)}\in\left(0.1\,{\rm dex}, 2.0\,{\rm dex}\right)$; $\log_{10} L_{\rm ref}^\mathrm{(QSO,Gal)}/\ergs\in\left(43.0,48.6 \right)$; $\gamma^\mathrm{(QSO,Gal)}\in\left(1,4 \right)$; 
$\log_{10} f_{\rm on}^\mathrm{(QSO,Gal)}\in\left(-3,0\right)$. 
The lower limits on $\sigma^\mathrm{(QSO,Gal)}$ and the upper limits $\log_{10} f_{\rm on}^\mathrm{(QSO,Gal)}$ are chosen because of physical constraints (i.e., the scatter in the $L-M$ relation is unlikely to be smaller than $0.1~\mathrm{dex}$ and the active fraction is less than unity by definition).

We provide joint constraints on the parameters by fitting the data described in Sec. \ref{sec:observations} (i.e., the luminosity and correlation functions for quasars and galaxies) simultaneously. In other words, we write the joint likelihood distribution as the product of the single Gaussian likelihoods for each dataset (we assume that all the measurements are independent):
\begin{equation}
    \mathcal{L}^\mathrm{(joint)} = \prod_i \mathcal{L}^\mathrm{(i)},
    \label{eq:likelihood_joint} 
\end{equation}
where $i$ ranges over the datasets shown in Table \ref{tab:obs_data}. 

When performing our analysis, we found that the data for the quasar auto-correlation function \citep[][]{arita2023}{}{} were not able to place significant constraints on any of our model parameters. As a result, this dataset was not informative, and could not be used to infer any of the physical properties of quasars. This conclusion differs from the one found in \citet[][]{arita2023}{}{}, where the authors are able to determine the range of host-halo masses for quasars at $z\approx6$. We investigated the issue further and found that the different conclusions arise from different assumptions made for the shape of the auto-correlation functions at large scales. For this reason, we exclude the \citet[][]{arita2023}{}{} dataset from the joint fit performed here, and defer the analysis of this dataset to Appendix \ref{sec:autos_arita}. In that Section, we compare in detail our analysis with the one performed by \citet{arita2023} and conclude that, if we assume a physically-motivated choice for the shape of the quasar auto-correlation function, we are not able to place interesting constraints on the distribution of quasar-host halo masses.

Moving forward, we discuss the results of our parameter inference for the ``joint'' model described above, including all the other datasets compiled in Table \ref{tab:obs_data}. We explore the posterior distribution for this model using a Markov-Chain Monte Carlo (MCMC) approach. We employ the Python package \code{emcee} \citep{foreman2013emcee} to sample the posteriors using the affine-invariant ensemble prescription \citep{goodman_prescription}. We place $m=48$ walkers distributed randomly in the parameter space and evolve them for $N=10^5$ steps. 
Figure \ref{fig:overview} (left panel) shows the corner plot for the 8-d posterior distribution, while Table \ref{tab:mcmc_results} summarizes the constraints we obtain for each of the model parameters. 
The samples of the posterior distribution obtained by the Markov Chains are then used to obtain predictions for the luminosity and correlation functions, both for quasars and galaxies at the same time; we compare these quantities with the data in the right panels of Figure \ref{fig:overview}.
The top right panel shows predictions for the galaxy luminosity function (orange) and the quasar luminosity function (red), while the bottom panel shows the quasar-galaxy cross-correlation function (red) and the galaxy auto-correlation function (orange). 

In all cases, we see that our model fares well when compared to the observational data. 
As a quantitative estimate of this accordance, we take the parameters corresponding to the maximum of the posterior distribution (highlighted by star symbols in the corner plot of Figure \ref{fig:overview}) and measure the $\chi^2$ statistic
for each of the single dataset shown in Table \ref{tab:obs_data}. Values of the $\chi^2$ are reported in the last column of Table \ref{tab:obs_data}. We generally find a very good agreement between our model and every single dataset analyzed. The only exception is the galaxy auto-correlation function, for which the $\chi^2$ is relatively large when compared to the size of the dataset. 
We believe this is due to the small reported uncertainties in the observational data, that are likely underestimated. As discussed in \citeE24, these uncertainties are assigned according to the Poissonian statistics associated with the pair counts, and they do not take into account the uncertainty coming from cosmic variance as well as other possible systematic effects. 
This may be particularly relevant in the outermost bins, for which the data drop significantly more rapidly than what is predicted by our model. Covariance between different data points is also neglected in the \citeE24 analysis, even though it most likely contributes to the total error budget significantly. This artificially increases the discrepancy between our model and the data.

\section{Results} \label{sec:results}

\begin{figure*}
	\centering
	\includegraphics[width=\textwidth]{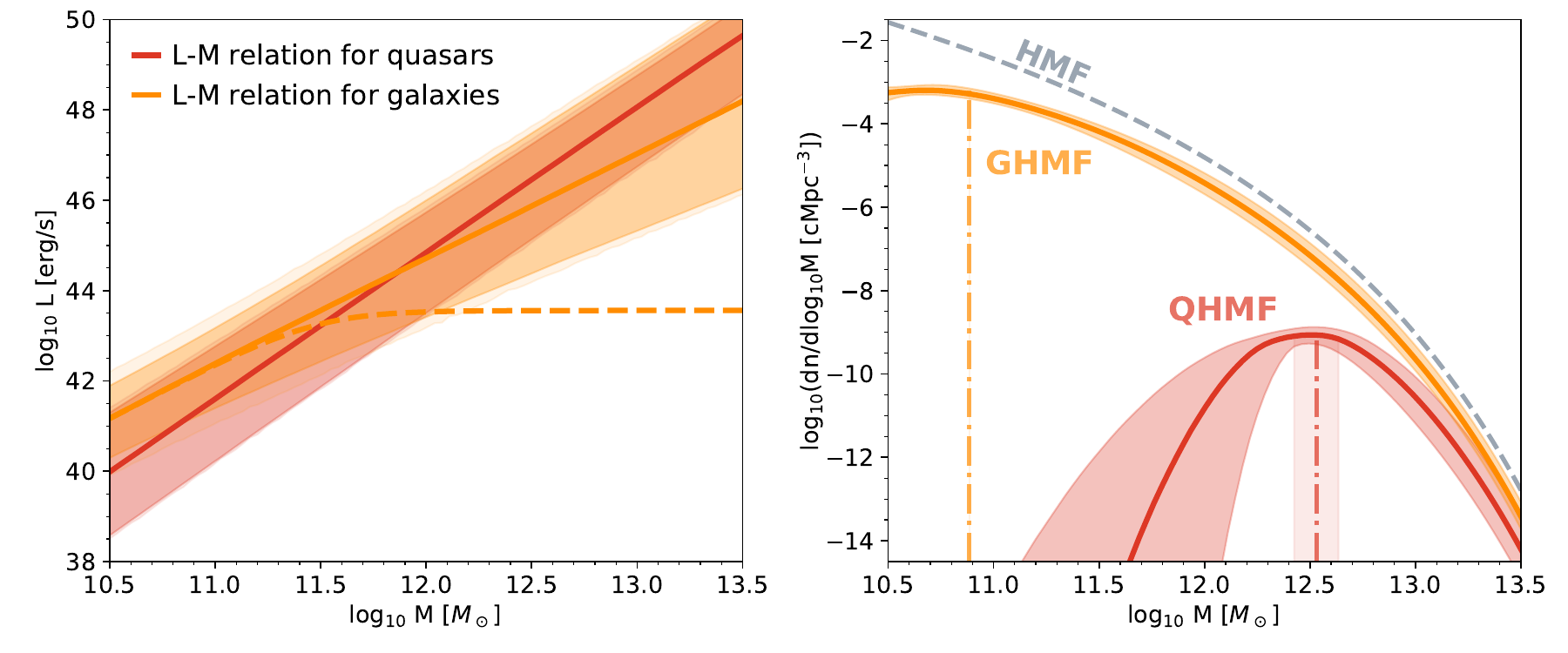}

	 \caption{{\it Left}: Luminosity-halo mass relation for quasars (red) and galaxies (i.e., \OIII emitters; orange). The quasar luminosity is the bolometric one, while the galaxy luminosity is the one from the \OIII line. Median values (solid lines) and 16th-84th percentiles (dark-shaded areas) for the $L-M$ relations are obtained by randomly sampling the Markov chains for the posterior distribution $2000$ times. The cumulative effects of the uncertainty on the median and the intrinsic scatter in the relations, expressed by the $\sigma$ parameter in the CLF, are shown with a lighter shading. The dashed orange line corresponds to the modified galaxy luminosity-halo mass relation, with a flattening of the relation above a threshold mass of $M=10^{11.5}\,\msun$.
     {\it Right}: Quasar-host mass function (QHMF; red) and galaxy-host mass function (GHMF; orange). Median and $1\sigma$ uncertainties of these functions (obtained as in the left panel) are shown with solid lines and dark-shaded areas, respectively. The dashed-dotted lines show the median halo masses associated with the QHMF (red) and GHMF (orange) distributions (see eq. \ref{eq:qhmf_med}); light-shaded regions represent $1\sigma$ uncertainties on these median masses. The halo mass function (HMF) at the redshift of interest ($z=6.14$) is shown with a gray dashed line.
    \label{fig:results_qg}
 	}
\end{figure*}

In the last Section, we have shown that we can successfully reproduce the data for the luminosity and correlation functions of quasars and galaxies with the simple extension of the \citeP24 framework described in Sec. \ref{sec:population_model}. In this framework, we use observations to constrain the conditional luminosity functions of quasars and galaxies simultaneously. In turn, each conditional luminosity function can be related to other fundamental properties such as the luminosity-halo mass relation, the host mass function, and the duty cycle/occupation fraction. We examine here these properties starting from the inferred values of the parameters obtained in Figure \ref{fig:overview} and Table \ref{tab:mcmc_results}. We first examine quasar properties, and then turn our attention to galaxies.

\subsection{The quasar luminosity-halo mass relation and the host halos of quasars at $z\approx6$}

Figure \ref{fig:results_qg} shows the quasar luminosity-halo mass relation (left) and the quasar-host mass function (QHMF; right) at $z\approx6$, as inferred from our model. We obtain a rather steep quasar $L-M$ relation, with a slope of $\gamma^\mathrm{(QSO)}\approx3.2$. This steep relation between quasar luminosities and halo masses is in agreement with the results of \citeP24, which use data at $z\approx2-4$ to study the evolution of this relation with redshift and find a significant increase in the slope parameter at earlier cosmic time. Our results suggest that this trend extends to even higher redshifts, with a close-to-linear relation at $z\approx2$ turning into a very steep relation ($\gamma^\mathrm{(QSO)}\approx2-3$) at $z\approx4-6$. We mention the caveat, however, that in this analysis the shape of the $L-M$ relation is primarily constrained by the QLF, and only marginally by the clustering measurements. This is because the \citeE24 clustering data only focus on a very bright sub-sample of $z\approx6$ quasars, and so they cannot constrain the behaviour of the $L-M$ relation below a luminosity of $\log_{10} L/\ergs \approx47$. Given that the shape and normalization of the QLF at high redshift are rather uncertain, 
especially at the faint end \citep[e.g.,][]{giallongo2019,Maiolino2023,harikane2023, andika2024}, the shape of the $L-M$ relation is inevitably also plagued by this uncertainty. 

The scatter in the quasar $L-M$ relation, on the other hand, is constrained both by the QLF and by the cross-correlation function simultaneously. In our analysis, we find a rather large log-normal scatter of $\sigma^\mathrm{(QSO)}\approx0.64$ dex (although with a significant uncertainty of $\approx0.3$ dex). This relatively large scatter is in line with the one measured by \citeP24 at $z\approx2.5$, but it represents a significant difference if compared to the very low scatter $\sigma^\mathrm{(QSO)}\lesssim0.3$ dex found by \citeP24 at $z\approx4$. Similarly, the value we obtain for the active fraction of $z\approx6$ quasars $f_\mathrm{on}^\mathrm{(QSO)}$ ($\approx2\%$) is rather low if compared to the very high active fraction ($\approx50\%$) found by \citeP24 at $z\approx4$. We defer the analysis of the peculiar redshift evolution traced by these parameters to Sec. \ref{sec:discussion_redshift}.

The QHMF (Figure \ref{fig:results_qg}, right panel) reveals that quasars tend to live in $\log_{10} M/\msun \approx 12.5$ halos (median value of $\log_{10} M/\msun \approx 12.53\pm0.13$), with a rather broad distribution encompassing a large range of halo masses (from $\log_{10} M/\msun \approx 12.1$ to $\log_{10} M/\msun \approx 12.8$ at $2\sigma$). The range of host masses we obtain is in perfect agreement with the conclusions of \citeE24, who pointed out that quasars tend to live in moderately strong over-densities, but not necessarily in the most over-dense regions of the Universe (corresponding to halo masses of $\log_{10} M/\msun\gtrsim 13$). 

Even more interestingly, the broad distribution of host masses that we find from the inferred QHMF is compatible with the large quasar-to-quasar variance in terms of over-densities found by \citeE24. The diversity of environments emerging from the \citeE24 observations is likely a combination of cosmic variance and variance in the host halo masses of quasars and/or galaxies. While we leave a quantitative analysis of these sources of variance to future work, it is encouraging to find evidence for the latter in our results. We stress the fact that our method for obtaining the QHMF does not make use of the observed diversity in terms of environments, as it only focuses on the global demographic and clustering properties of galaxies and quasars. The broad distribution of host masses that we find from our QHMF follows naturally from jointly modeling the clustering properties of quasars together with the shape and normalization of the quasar luminosity function. 

In the analysis presented in \citeE24, the framework developed here was used to match the quasar-galaxy cross-correlation function and the galaxy auto-correlation function by assuming simple ``step-function'' halo occupation distributions (HODs) for both quasars and galaxies. In other words, \citeE24 populated halos and galaxies only above some minimum mass thresholds. With this method, they inferred the \textit{minimum} host halo mass for quasars to be $\log_{10} M/\msun \approx 12.43$.  For a ``step-function'' HOD model, this value corresponds to a \textit{median} quasar host mass of $\log_{10} M/\msun \approx 12.51$, in excellent agreement with the median value of our inferred QHMF distribution.

Our conclusions on the quasar-host masses are also in line with the ones obtained by Mackenzie et al.\ in prep. In this work, the authors use the UniverseMachine \citep{universe_machine} to compare the number of satellite halos to the number of companion galaxies observed in EIGER quasar fields. In this way, they obtain a distribution of possible host dark matter halo masses for each observed quasar in \citeE24. Overall, the median value they obtain by putting together all the different mass distributions is $\log_{10}M/\msun=12.4\pm0.5$. The agreement with our results is significant, considering the very different assumptions underlying this method compared to the ones made here. Another estimate for the typical host halo masses of EIGER quasars was also obtained in \citeE24 by comparing the observed $\chi_{V,\mathrm{QG}}$ with predictions from the TRINITY model \citep{trinity}. The resulting median host halo mass, $\log_{10} M/M_\odot = 12.14 ^{+0.24}_{-0.26}$, is slightly lower than the one found here, but still marginally compatible when taking uncertainties into account.

Finally, by relating the inferred QHMF to the halo mass function (HMF) at the same redshift (see eq. \ref{eq:duty_cycle}), we can obtain an estimate for the quasar duty cycle, $\varepsilon_\mathrm{QDC}$. Fig. \ref{fig:results_dc} (left panel) shows the probability density function (PDF) for the quasar duty cycle (red) and the galaxy duty cycle (orange) obtained by randomly sampling the Markov chains for the posterior distribution shown in Fig. \ref{fig:overview}. We infer a value for the quasar duty cycle of $\varepsilon_\mathrm{QDC}=0.9^{+2.3}_{-0.7}\%$. This relatively low value of the duty cycle implies that only a small fraction of SMBHs 
are active as UV-bright, luminous quasars at any given time, and it has relevant consequences in terms of the lifetime of high-$z$ quasars, their obscuration fraction, and more generally the growth of SMBHs. We will explore this further in Sec. \ref{sec:dc_smbh_growth}.

\subsection{Characterizing the properties of [OIII] emitters}

\begin{figure*}
	\centering
	\includegraphics[width=\textwidth]{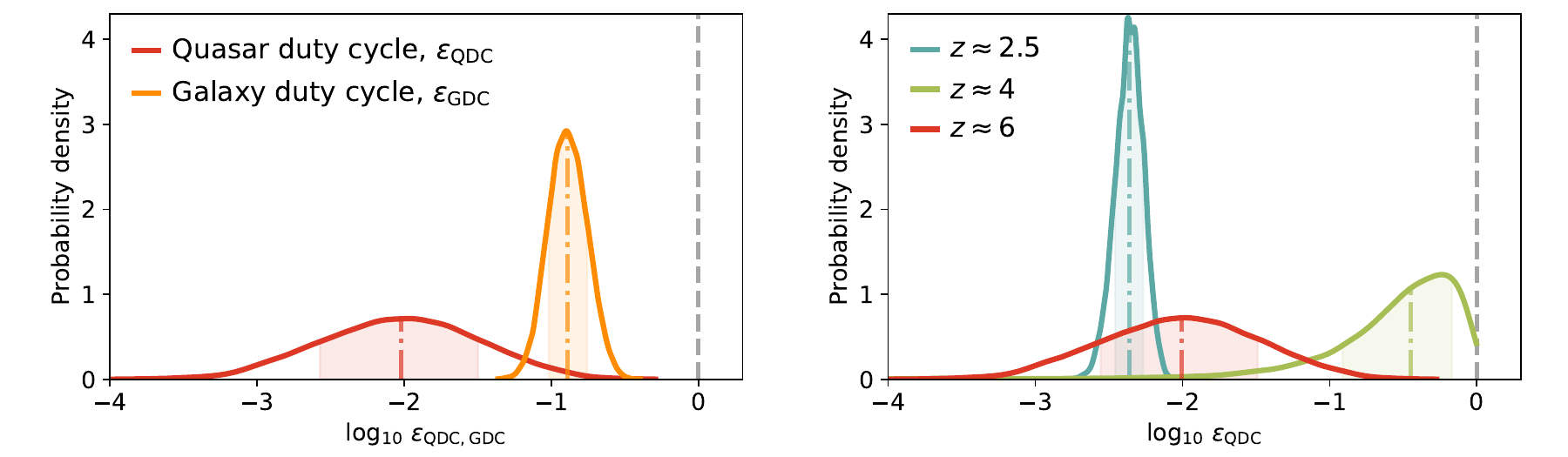}

	 \caption{{\it Left}: Probability density function (PDF) for the quasar (red) and galaxy (orange) duty cycles at $z\approx6$, obtained by randomly sampling the Markov chains for the posterior distribution $2000$ times. The median and $1\sigma$ uncertainties for the PDFs are shown with a dot-dashed line and shaded areas, respectively. The dashed vertical line corresponds to the maximum possible value of the duty cycle, $\varepsilon_\mathrm{DC}=100\%$.
     {\it Right}: Same as the left panel, but for the quasar duty cycles at different redshifts: $z\approx2.5$ (blue), $z\approx4$ (green), and $z\approx6$ (red). Results at redshift lower than $z\approx6$ are taken from \citeP24.
    \label{fig:results_dc}
 	}
\end{figure*}

Our joint model for quasars and galaxies constrains the properties of these two populations simultaneously. As a result, all the properties that we have presented for quasars can also be studied for the high-$z$ galaxy population. These are the galaxy luminosity-halo mass relation (Fig. \ref{fig:results_qg}, left panel), the GHMF (Fig. \ref{fig:results_qg}, right), and the galaxy duty cycle (Fig. \ref{fig:results_dc}, left panel). Before analyzing these quantities, we note that our model focuses only on \OIII emitters, as this sub-population of galaxies is the one that is targeted by the JWST NIRCam-WFSS observations from the EIGER survey. Therefore, all results that we will quote here refer to the properties of galaxies that are bright in the \OIII line; at these high redshifts, these galaxies are believed to be luminous, star-forming, and unobscured \citep[e.g.,][]{matthee23_eiger}{}{}. 

The galaxy luminosity-halo mass relation (Fig. \ref{fig:results_qg}, left panel) is rather similar to the quasar luminosity-halo mass relation. The major differences can be found in the slope of this relation as well as in its normalization. The logarithmic slope of the galaxy luminosity-halo mass relation is shallower than the one concerning quasars, but steeper than linear ($\gamma^\mathrm{(Gal)}\approx2.3$). The normalization of this relation conspires with its slope to give an average galaxy luminosity at fixed halo mass that is brighter than the one of quasars at $\log_{10} M/\msun \lesssim 11.5$, but dimmer at larger host halo masses\footnote{Note, however, that the luminosity of galaxies only includes the flux emitted in the \OIII line, hence we expect the normalization of the galaxy $L-M$ relation to be higher when considering the total flux emitted from galaxies.}. 
This implies that, on average, quasars overshine galaxies at the high mass end of the HMF, while the opposite is true for the bulk of the halo population.

Nonetheless, if we look at the comparison between the QLF and the GLF in Fig. \ref{fig:overview} (top right panel), we see that our model predicts galaxies to be more abundant than quasars at all luminosities. This is because the scatter in the galaxy $L-M$ relation is rather large, and the duty cycle of galaxies is significantly larger than that of quasars (see below). 
Observationally, we know that the GLF drops below the QLF at luminosities around $\log_{10} L/\ergs \approx 46$ \citep[e.g.,][]{bouwens2015, matsuoka2018}{}{}, so this implies that the extrapolation of the GLF at large luminosities based on our model is flawed. This is not a surprise, as here we assumed that a very simple power-law relation between galaxy luminosity and halo mass holds for the entire population of halos. This relation serves our purposes, as we want to match data for the GLF only in a rather narrow luminosity range, but it is probably too simplistic to capture the behaviour of the galaxy population at even larger luminosities/host masses. 

Indeed, we know that the star formation efficiency 
is predicted to peak for halo masses of $\log_{10} M/\msun \approx 11.5-12.5$, resulting in a break in the stellar mass-halo mass relation \citep[e.g.,][]{moster2013,universe_machine}{}{}. While the luminosity range of the GLF data considered here is not large enough to constrain this break in the context of our model, we can see what would be the effect of a more physically motivated choice for the galaxy $L-M$ by making the arbitrary assumption that this relation flattens above $\log_{10} M/\msun\approx11.5$ (dashed line in the left panel of Fig. \ref{fig:results_qg}). In practice, we assume that the galaxy CLF in eq. \ref{eq:clf_log_normal} remains the same, but we vary the luminosity-mass relation on which it is based (eq. \ref{eq:power_law_Lc}) by manually inserting a flattening above a threshold halo mass. We find that all the quantities but the GLF remain unchanged; the new median GLF is plotted with a dashed line in Fig. \ref{fig:overview} (top right panel). Indeed, we see that with this simple assumption, the predicted GLF drops below the QLF at roughly the observed luminosity. A more comprehensive quasar/galaxy population model -- that is outside the scope of this paper -- would include a larger set of galaxy observations to properly constrain the shape of the break in the galaxy $L-M$ relation. The simple argument adopted here, however, shows that our framework is well-suited to represent quasars and galaxies in the luminosity/mass ranges of the data we aim to reproduce (Table \ref{tab:obs_data}).

The GHMF is shown in the right panel of Figure \ref{fig:results_qg} (orange line). Again, we find a broad distribution of host masses, with a median value of $\log_{10} M/\msun \approx 10.9$ ($\log_{10} M/\msun= 10.88^{+0.04}_{-0.03}$) and a $1\sigma$ range of $\pm 0.3$. 
Determining the characteristic host halo masses for \OIII emitters is an important result that is made possible by the analysis presented here. This population of galaxies is a major protagonist in JWST campaigns to study the high-$z$ Universe via slitless spectroscopy \citep[][]{kashino2022, oesch2023_fresco,aspire_wang2023}{}{}. For this reason, a thorough characterization of their properties is key. Overall, the characteristic host mass that we find for \OIII emitters agrees well with the one measured at the same redshifts using Lyman break galaxies (LBGs) in HST photometric campaigns \citep[][]{barone-nugent2014, dalmasso2024}{}{}. This result strengthens the conclusion 
 -- coming from abundance arguments \citep[][]{matthee23_eiger}{}{} -- that \OIII emitters may trace star-forming regions in high-$z$ galaxies in a way that is similar to Lyman-break-selected systems.   

We note that the shape of the GHMF (Fig. \ref{fig:results_qg}, right panel) is affected by the minimum mass we assume in our model, i.e., $\log_{10} M_\mathrm{min} /\msun = 10.5$ (see Sec. \ref{sec:methods}). In other words, in our population model, we assume that galaxies live only in halos larger than this threshold mass, and that the GHMF goes to zero for lower masses. This choice is made in the context of our framework because the FLAMINGO-10k simulation introduced in Sec. \ref{sec:simulations} cannot resolve halos with lower masses. There is no physical motivation, however, for this choice, as there could be a population of bright galaxies that are residing in lower-mass halos. In particular, we believe that extending the GHMF distribution to lower halo masses would bring the median value found here ($\log_{10} M/\msun \approx10.9$) down to slightly lower values. This is because the GHMF distribution is artificially skewed towards larger halo masses because of the halo mass threshold imposed in our simulation: the halo mass corresponding to the peak of the GMHF distribution ($\log_{10} M/\msun \approx10.7$) is lower than the median ($\log_{10} M/\msun \approx10.9$). Indeed, a lower median value of $\log_{10} M/\msun \approx10.7$ is in closer agreement with the result found in \citeE24, where the same simulation presented here was coupled with a ``step-function'' HOD model for quasars and galaxies. The authors found a \textit{minimum} host mass for \OIII emitters of $\log_{10} M/\msun \approx 10.56$, which can be translated into a median mass of $\log_{10} M/\msun \approx 10.65$. Nonetheless, we believe that extending the model to lower halo masses would not significantly impact the conclusions presented here: we experimented with different prescriptions for the GHMF and always found similar results, with the median value of the GHMF of $\log_{10} M/\msun \approx10.8-10.9$) and the peak of the GHMF distribution at $\log_{10} M/\msun \approx10.6-10.7$. 
Using a simulation with a smaller volume and higher resolution, one could resolve halos down to much lower masses and hence fully capture the properties of galaxies and their host halos. However, this is not the goal of our work, as the primary focus of our analysis is the relation between quasars and the galaxies in their environments, which can only be captured with a large-volume simulation given the rarity of quasars at high-$z$.

The galaxy duty cycle, $\varepsilon_\mathrm{GDC}$, is a measure of how many halos host galaxies that can be observed in \OIII compared to the global halo population with the same characteristic masses. In our model, we infer a value for the galaxy duty cycle of $\varepsilon_\mathrm{GDC}=12.9^{+4.7}_{-3.3}\%$. This is once again in agreement with the duty cycle values inferred from LBG clustering analysis \citep[e.g.,][]{dalmasso2024}{}{}. We note here that the notion of ``duty cycle'' is primarily utilized in the context of quasars rather than galaxies, as gas accretion on SMBHs -- that is believed to be associated with the triggering of quasar activity -- is assumed to be episodic, and hence the whole process is cyclic in cosmic time. In the context of galaxies, it is probably easier to talk about an ``occupation fraction'' of \OIII emitters, implying that only a fraction of halos is hosting galaxies whose \OIII emissions are bright enough to be detectable and not obscured by dust. However, it is also relevant to point out that if \OIII emitters, as argued before, trace unobscured star formation, they may also be subject to rapid change in their luminosity as the star formation process is also thought to be episodic, especially at high redshifts \citep[e.g.,][]{faucher-giguere2018,pallottini2023}{}{}. Indeed, UV-variability \citep[e.g.,][]{shen2023, sun2023}{}{} has been argued to play a key role in explaining the over-abundance of bright galaxies that was indicated by JWST imaging at very high-$z$ \citep[e.g.,][]{naidu2022, finkelstein2023}{}{}. 

Our duty cycle measurement cannot determine the amount of variability in the galaxy lightcurves, as it only offers an integral constraint on the total light emitted (in the \OIII line) by star-forming galaxies over the entire history of the Universe. In other words, it is only sensitive to the zeroth moment of the galaxy's unobscured lightcurve distribution. 
Nonetheless, the value of the duty cycle inferred here represents an important independent characterization of the star formation history of high-$z$ galaxies, and it nicely complements probes of the burstiness of the high-$z$ star formation process coming from spectral energy distribution (SED) fitting \citep[e.g.,][]{looser2023,endsley2023,cole2023}{}{}.

Another interesting point to make here is that the duty cycle/occupation fraction that we measure for galaxies sets an upper limit on the contribution of obscured star formation to the total galaxy mass growth at early times. This is because our measurements tell us that $\gtrsim15\%$ of $z\approx6$ galaxies are \OIII-bright, and hence the fraction for which star formation is obscured by dust cannot be higher than $\approx85\%$. 
This is an interesting constraint that can be directly compared with the estimated fraction of obscured star formation coming from ALMA observations \citep[e.g.,][]{algera2023}{}{}. We will return to the point of obscuration in the context of quasars and SMBH growth in Sec. \ref{sec:dc_smbh_growth}

\section{Discussion} \label{sec:discussion}

\begin{figure*}
	\centering
	\includegraphics[width=\textwidth]{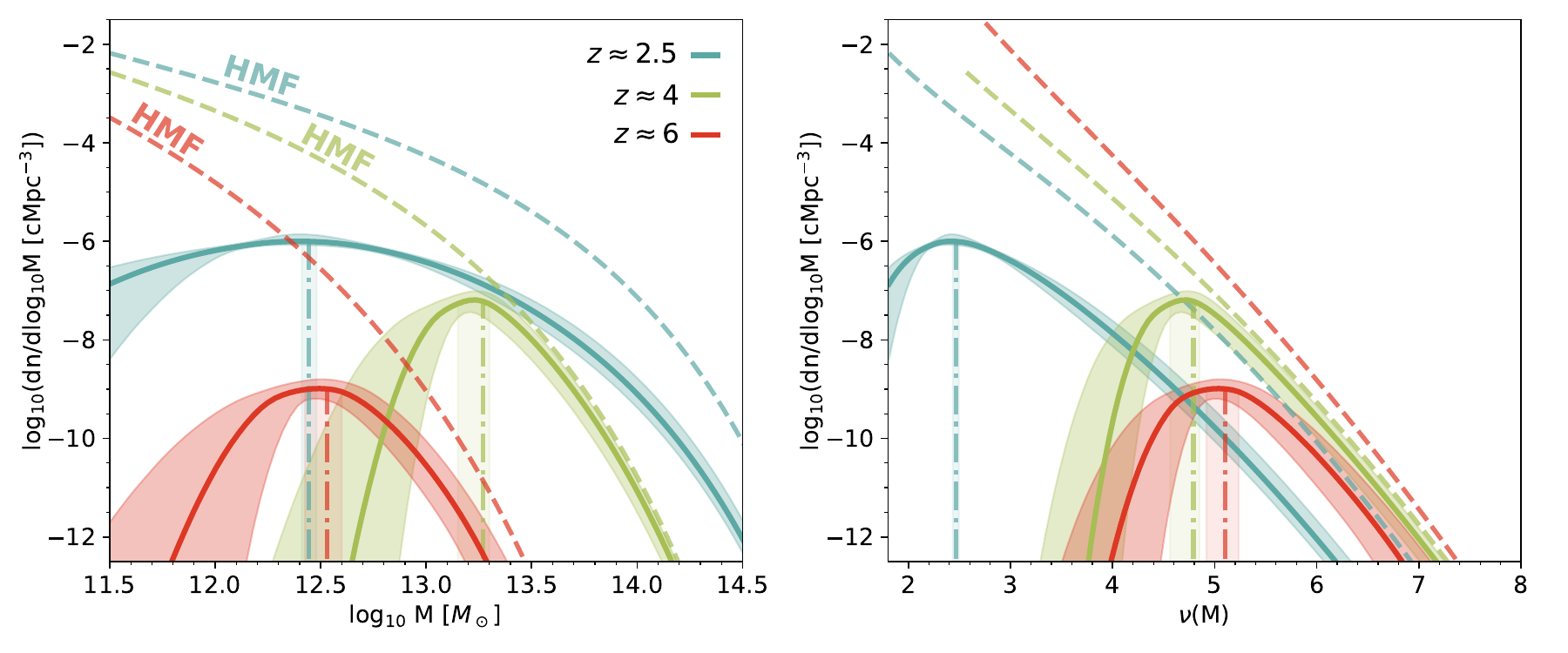}

	 \caption{{\it Left}: Quasar-host mass function (QHMF) at $z\approx2.5$ (solid blue line), $z\approx4$ (solid green line), and $z\approx6$ (solid red line) as a function of the halo mass, $M$. The halo mass functions (HMFs) at the same redshifts are shown with dashed lines and color-coded in the same way as the QHMFs. The dashed-dotted lines represent the median values of the QHMF distributions (see eq. \ref{eq:qhmf_med}), while shaded regions represent $1\sigma$ uncertainties on the various quantities.  {\it Right}: Quasar-host mass functions (QHMFs; solid lines) and halo mass functions (HMFs; dashed lines) as a function of the peak height, $\nu(M)$, at different redshifts. Color codes and other quantities are the same as in the left panel.   
  \label{fig:results_z}
 	}
\end{figure*}

In the analysis performed above, we could successfully match the luminosity functions and the clustering properties of quasars and galaxies at $z\approx6$ provided that: (a) there exist non-linear relations between quasar/galaxy luminosity and halo mass; (b) these relations have significant log-normal scatter ($\sigma \approx 0.5-1$ dex), and the one for quasars is steeper ($\gamma^\mathrm{(QSO)}\approx3.2$) than the one for galaxies ($\gamma^\mathrm{(Gal)}\approx2.3$); (c) following these relations, luminous quasars ($\log_{10} L/\ergs \gtrsim 47$) are hosted by halos with mass $\log_{10} M/\msun \approx 12.5$, while galaxies ($\log_{10} L/\ergs \gtrsim 42.5$) are hosted by much smaller halos with $\log_{10} M/\msun \approx 10.9$; (d) (UV-bright) quasars occupy only a small fraction of halos with a duty cycle $\varepsilon_{\rm QDC}\approx0.9\%$, while the occupation fraction/duty cycle of galaxies is significantly larger, $\varepsilon_{\rm GDC}\approx13\%$.

In the following, we further elaborate on this picture by focusing on the properties of high-$z$ quasars, studying their implications for SMBH accretion and growth and their evolution with cosmic time. 

\subsection{Quasar properties across cosmic time}
\label{sec:discussion_redshift}

In \citeP24, we applied a very similar framework to the one presented here to model the auto-correlation and luminosity functions of quasars at $z\approx2.5$ and $z\approx4$. As a result, we obtained the quasar luminosity-halo mass relation, the QHMF, and the quasar duty cycle at these two different redshifts, and discussed how the properties of quasars seem to evolve rapidly between these two epochs. Thanks to the analysis performed here, we can now extend this discussion to include the properties of $z\approx6$ quasars, and attempt to paint a coherent picture of quasar evolution in the first few billion years of the Universe. The right panel of Fig. \ref{fig:results_dc} shows the PDFs for the inferred values of the quasar duty cycles at $z\approx2.5$ (blue), $z\approx4$ (green), and $z\approx6$ (red). The first two curves are obtained by sampling the posterior distributions for the parameters from \citeP24 (see their Fig. 5), while the last one is the same as in the left panel. 
The same plot but for the QHMF is shown in the left panel of Fig. \ref{fig:results_z}. 

Quite interestingly, we see that the evolution of the QHMF and the quasar duty cycle with redshift do not follow a monotonic trend. The duty cycle is low ($\lesssim0.5\%$) at $z\approx2.5$, but it increases rapidly to values $\gtrsim50\%$ at $z\approx4$. At even higher redshifts, however, the duty cycle seems to drop again to $\lesssim1\%$. Despite the relatively large uncertainty on our $z\approx6$ measurement, the difference with the result obtained at $z\approx4$ is rather remarkable (Fig. \ref{fig:results_dc}, right panel). An analogous trend with redshift can be observed by considering the median of the QHMF distribution, which represents the characteristic mass for the population of halos that are hosting bright quasars (Fig. \ref{fig:results_z}, left panel): this median mass is $\approx10^{12.4-12.5}\,\msun$ for $z\approx2.5$ and $z\approx6$, while it grows to $\approx10^{13.3}\,\msun$ at $z\approx4$. 

As discussed in \citeP24, the rather extreme values of the duty cycle and the host masses that we find at $z\approx4$ are a consequence of the very strong quasar clustering measured by \citet[][]{shen2007}{}{}. Using data from the Sloan Digital Sky Survey (SDSS), \citet[][]{shen2007}{}{} find a value of the quasar auto-correlation length, $r_{0,\mathrm{QQ}}$, of $\approx24\,\cMpch$, which is significantly higher than the value $r_{0,\mathrm{QQ}}\approx8\,\cMpch$ measured by \citet[][]{eftekharzadeh2015}{}{} \citep[see also][]{ross2009,shen2009,white2012}{}{} at $z\approx2-3$ using the BOSS survey. The strong quasar clustering at $z\approx4$, combined with a rather large abundance of bright quasars at the same redshift ($\approx3\times10^{8}\,\cMpc^{-3}$), implies that only very massive halos can host active SMBHs and a large fraction of them are continuously shining as quasars at any given moment (i.e., the quasar duty cycle is large). This -- as discussed by several works (\citeP24,\citealt{white_2008,shankar_2010}) -- is only possible provided that the scatter in the relation between quasar luminosity and halo mass is very low ($\sigma\lesssim0.3$ dex). 

The analysis presented here shows that the trend hinted by the \citet[][]{shen2007}{}{} quasar clustering measurements at $z\approx4$ does not seem to extend further to higher redshifts. Using the data from \citeE24, we have shown that the characteristic host mass of quasars at $z\approx6$ is not as large, and only a small fraction of these SMBH-hosting halos are actually shining as bright quasars at any given time. As a consequence, the tight constraints on the scatter between quasar luminosity and halo mass are not in place at $z\approx6$, and our model finds a larger value for the scatter of $\sigma\approx0.6$ dex, although lower values are also compatible with the data (Fig. \ref{fig:overview}). Overall, these results may suggest that the measurements of quasar clustering at $z\approx4$ \citep[][]{shen2007}{}{} may be overestimated \citep[see also][]{he2018, de_beer2023}{}{}, and that the constraints on the host masses, quasar duty cycle, and scatter in the $L-M$ relation at $z\approx4$ may need to be relaxed to some extent. If that is the case, our results at $z\approx6$ suggest that quasars are hosted, on average, by a small fraction of the population of halos with masses in the range $\approx10^{12}-10^{13}\,\msun$, in line with the situation at $z\approx2-3$. This result may favor a picture in which there exists a range of halo masses for which quasar activity can be supported that is independent of cosmic time. According to this picture, halos whose masses are lower than this range cannot be responsible for a significant fraction of the black holes that are capable of turning into bright quasars, while for very massive hosts ($\log_{10}M/\msun\approx13$) quasar activity is quenched by feedback mechanisms \citep[e.g.,][]{hopkins2007,fanidakis2013,caplar2015}{}{}. 

On the other hand, the measurements from \citet[][]{shen2007}{}{} appear to be solid, 
being based on a large ($\approx5000$) spectroscopic sample of high-$z$ quasars from SDSS, and they are also backed up by estimates of the small-scale quasar clustering inferred from 
independent samples of $z\approx4-5$ binary quasars \citep[][]{hennawi2010,mcgreer2016,yue2021}{}{}. It is thus worth taking the \citet[][]{shen2007}{}{} clustering data at face value, and exploring the implications of their results in terms of the evolution of quasar properties at early cosmic times. The \citet{shen2007} measurements suggest that at high redshifts quasar activity tends to take place only in the most massive halos, tracking halo growth across cosmic time \citep[][]{hopkins2007}{}{}. It is interesting to note that our $z\approx6$ results do not necessarily disfavor this scenario. In fact, our inferred QHMFs suggest that quasars live \textit{in equivalent halos} at $z\approx4$ and $z\approx6$, while they live in very different environments at lower redshifts. This can be understood by looking at the right panel of Fig. \ref{fig:results_z}, which shows the QHMFs at different redshifts plotted as a function of the peak height, $\nu(M)$. The peak height is defined as $\nu(M,z) = \delta_c/\sigma(M,z)$ -- with $\delta_c\approx 1.69$ being the critical linear density for 
spherical collapse and $\sigma^2(M,z)$ the variance of the linear density field smoothed on a scale $R(M)$
\footnote{We compute $\nu(M,z)$ using the python package \code{colossus} \citep[][]{colossus_diemer2018} and setting the same cosmology as the FLAMINGO-10k simulation (Sec. \ref{sec:simulations}). However, we mention the caveat that the definition of peak heights implicitly assumes that halo masses are based on the spherical overdensity formalism, and it only applies to the current masses of central halos (and not to satellites). In our analysis (Sec. \ref{sec:hbt}), we assume a halo mass definition based on peak bound masses instead, and include the contribution of satellites as well. Nonetheless, we believe that the effects of the differences in halo mass definition are relatively small and that the final values we obtain for the peak heights are not impacted significantly by these factors.}. 
It is a way to relate the masses of halos at any redshifts to the strength of the fluctuations in the initial conditions of the original linear density field. 
Therefore, large (small) peak heights correspond to very over(under)-dense environments, independently of redshift.

The right panel of Fig. \ref{fig:results_z} shows that quasars tend to be hosted by very rare $\approx5\sigma$ fluctuations both at $z\approx6$ and $z\approx4$. This suggests that the same kind of rare and very biased halos host bright quasars at early cosmic times, and that these host halos are more massive at $z\approx4$ than at $z\approx6$ only because they grow via mergers/accretion during the $\approx700$ million years of cosmic time that separate these two redshifts. 
In the lower redshift Universe ($z\approx0-3$), instead, the situation is quite different, with quasars being hosted by a new, less biased population of halos which corresponds to $\lesssim3\sigma$ fluctuations in the density field.

In this scenario, the key difference between $z\approx6$ and $z\approx4$ is the \textit{duty cycle}: while at $z\approx4$ almost all of the most massive halos need to host UV-bright quasars, the fraction of these same halos that are revealed as quasar hosts at $z\approx6$ is dramatically smaller. This could be caused by either much shorter and sparser accretion episodes at very early cosmic times or a much larger obscuration fraction characterizing early SMBH accretion. It is of great interest to relate these arguments to our current paradigm of SMBH growth: this will be the subject of Sec. \ref{sec:dc_smbh_growth}.

In order to discriminate between the scenarios discussed here and to paint a complete evolution of quasar activity across cosmic time, it is essential to investigate the clustering of quasars at high redshifts with new methods and new observational campaigns. In this sense, the next few years promise to bring a new wealth of data with the combined action of JWST mapping quasar-galaxy clustering at different redshifts using NIRCam WFSS \citep[][]{kashino2022,aspire_wang2023}{}{}, and the DESI survey \citep[][]{DESI}{}{} using ground-based spectroscopy to unveil a new, large sample of quasars up to $z\lesssim5$ that can be used to compute the quasar auto-correlation function with a much higher precision.

We conclude by mentioning the caveat that the QHMFs shown in Fig. \ref{fig:results_z} are obtained by setting luminosity thresholds that vary according to the ones used in observational data. In other words, the definition of ``bright'' quasars we employ is redshift-dependent, and it is based on the depth of the survey that was used for the clustering measurements. In Appendix \ref{sec:same_Lthr}, we show the same QHMFs obtained by setting a uniform luminosity threshold of $\log_{10} L_\mathrm{thr}/\ergs=46.7$, which is the same luminosity threshold as used by \citet[][]{shen2007}{}{} at $z\approx4$ and roughly corresponds to the break of the quasar luminosity function at all redshifts $z\gtrsim2$ \citep[e.g.,][]{khaire2015, kulkarni2019}{}{}. The resulting QHMF shifts towards higher (lower) halos masses at $z\approx2.5$ ($z\approx6$), due to the different luminosity thresholds employed in observations with respect to the one at $z\approx4$. Nonetheless, the global picture that we presented in this Section remains unchanged: quasars seem to be hosted by $\log_{10} M/\msun \gtrsim 13-13.5$ halos only at $z\approx4$, but when relating halo masses to their large-scale environments by using the peak height formalism, we find a direct connection between $z\approx4$ and $z\approx6$ and a divergent behavior at lower redshifts.

\subsection{The quasar duty cycle and SMBH growth}\label{sec:dc_smbh_growth}

One of the key results of our analysis is that the quasar duty cycle we obtain at $z\approx6$ is rather low ($\approx0.9\%$), in stark contrast with the very high one ($\gtrsim50\%$) measured at $z\approx4$ from the \citet[][]{shen2007}{}{} data (Fig. \ref{fig:results_dc}, right panel). As detailed in, e.g., \citeP24, these duty cycles can be directly converted into estimates of the total time SMBHs shine as bright quasars (i.e., the integrated quasar lifetime, $t_\mathrm{Q}$) via the simple relation $t_\mathrm{Q}=t_\mathrm{U}(z)~\varepsilon_\mathrm{DC}$ -- with $t_\mathrm{U}(z)$ being the age of the Universe at a given redshift.
Using the values of the duty cycles mentioned above, we obtain {$t_\mathrm{Q}\approx0.1-1~\mathrm{Gyr}$ at $z\approx4$, and a smaller $t_\mathrm{Q}\approx10~\mathrm{Myr}$ at $z\approx6$.
It is important to investigate the discrepancy between the values obtained at these two redshifts further, as the study of the timescales of quasar activity at high redshift is intrinsically connected with the formation and evolution of SMBHs in the Universe. 

As discussed in the Introduction, our current paradigm of SMBH growth is founded on the idea that SMBHs grow by accretion, and that a small fraction of the accreted rest mass is converted into radiation and gives rise to the quasar phenomenon. According to this paradigm, the growth of SMBHs is always concomitant with the formation of a bright quasar. For this reason, the total time a SMBH shines as a quasar (i.e., the quasar lifetime) is related to the total mass that has been accreted onto the SMBH. This argument has been proposed in many different variations in the past \citep[e.g.,][]{soltan1982, martini2001, yu_tremaine2002}{}{}, and it represents one of the cornerstones of our understanding of quasar/SMBH evolution. 

At high redshift ($z\gtrsim6$), the connection between 
the quasar lifetime and SMBH growth is even more relevant 
due to the limited amount of cosmic time ($\lesssim1\,\mathrm{Gyr}$) 
that is available to grow black holes to the observed masses of $\approx10^{8-9}\,\msun$ \citep[][]{fan2022}{}{}. Assuming Eddington-limited growth with a standard radiative efficiency of $\approx10\%$, one finds that only by postulating $t_{\rm Q}\sim 0.1-1\,\mathrm{Gyr}$ (i.e., a quasar duty cycle $\gtrsim10\%$) it is possible to explain the presence of such black holes in the early Universe starting from massive black hole seeds of $\approx10^{3-5}\,\msun$ \citep[e.g.,][]{inayoshi2020, pacucci2022}{}{}. This argument agrees well with the long lifetime inferred by our model at $z\approx4$ (see \citeP24 for further discussion), but it is in plain tension with the low duty cycle at $z\approx6$ that we inferred in this work. 

This tension between the long timescales required by SMBH growth and the short timescales that seem to be associated with high-$z$ quasar activity has already been investigated in the context of quasar proximity zones and damping wing features. By looking at quasar rest-frame UV spectra, several studies at $z\approx4-7$ have argued that the inferred quasar lifetimes range between $t_\mathrm{Q}\approx0.1-10\,\mathrm{Myr}$ \citep[e.g.,][]{khrykin2016, khrykin2019,khrykin2019b, Eilers2018b, Eilers2020, davies2018,Davies2019b,davies2020,worseck2016,worseck2021,Durovcikova2024}, 
and do not seem to reach the very large values required by SMBH growth models. Constraints based on proximity zones/damping wings are sensitive to the local conditions of each quasar environment 
and only probe a fraction of the past quasar lightcurve, so the direct connection between these results and the ones related to quasar clustering -- which probe the global population of quasars and can only constrain their total lifetime -- is non-trivial in the presence of rapidly varying and/or flickering lightcurves \citep[e.g.,][]{davies2020,satyavolu2023}.

Nonetheless, the cumulative evidence coming from these very different probes of quasar activity indicates that our standard paradigm for SMBH growth at high $z$ may need to be thoroughly reconsidered: not only is there very little cosmic time to grow the SMBHs to the billion solar masses that we observe for bright $z\approx6-8$ quasars, we also lack evidence for this accretion taking place in the form of UV-bright quasar emission at $z\gtrsim6$. 
Proposed solutions to this problem include a very low radiative efficiency $\lesssim0.1-1\%$ -- which implies that only a very small fraction of the accreted mass is converted into quasar light -- or a very large population of obscured SMBHs at high-$z$ that are not visible as UV-bright quasars but continue to grow actively at all epochs \citep[e.g.,][]{Davies2019b}. 
This latter hypothesis is particularly relevant, as a large obscured fraction for $z\gtrsim6$ quasars has been proposed both in the context of cosmological simulations \citep[e.g.,][]{Ni2020, Vito2022, Bennett2024} and multi-wavelength observations 
\citep{Vito18,Circosta19,DAmato20,Gilli2022, endsley2023}. Recently, JWST data have unveiled a new population of candidate dust-obscured active galactic nuclei (AGN) that can only be found at high redshifts \citep{harikane2023, Matthee2023b, kocevski2023, Maiolino2023, Greene2023, Kokorev2023, Kokorev2024, lin_aspire2024}, and may suggest a rapid evolution of the obscuration properties of AGN/quasars in the early Universe.

\section{Summary} \label{sec:conclusions}

In this work, we have modeled the demographic and clustering properties of quasars (i.e., type-I, UV-bright systems) and galaxies (i.e., \OIII emitters) at $z\approx6$ using an extension of the framework introduced in \citet[][]{pizzati2023}{}{} (\citeP24; see their Figure 1). The model presented here builds on a new, state-of-the-art N-body simulation from the FLAMINGO suite \citep[][]{flamingo}{} (the ``FLAMINGO-10k'' run) that has the same resolution as the original FLAMINGO DMO high-resolution run (CDM particle mass of $8.40\times10^{8}\,\msun$) but a much larger volume ($L=2.8\,\cGpc$). 

Thanks to this simulation, we can model the properties of $z\approx6$ quasars and galaxies simultaneously; these include (Table \ref{tab:obs_data}): the galaxy luminosity function (\citealt{matthee23_eiger}), the quasar luminosity function \citep{schindler2023}, the quasar-galaxy cross-correlation function and the galaxy auto-correlation function \citep[][]{eilers2024}{}{}, and the quasar auto-correlation function \citep[][considered separately in Appendix \ref{sec:autos_arita}]{arita2023}{}{}. 

The model we employ is founded on a Conditional Luminosity Function (CLF) framework. We assume a CLF for both quasars and galaxies, with identical parameterizations, i.e., power-law relations between quasar/galaxy luminosity and halo mass ($L\propto M^\gamma$) with log-normal scatter $\sigma$. We also include an active fraction, $f_\mathrm{on}$, to account for the fraction of quasars/galaxies that are too dim or not active and hence cannot be detected by observations.

The CLFs effectively link the population of halos in the simulated volume to the ones of quasars/galaxies. Therefore, once the halo mass function is known, we can directly obtain the quasar/galaxy luminosity function and the quasar-/galaxy-host mass function (QHMF/GHMF). The QHMF/GHMF can be coupled to the cross-correlation functions of halos with different masses to model the clustering properties (auto-/cross-correlations) of quasars and galaxies simultaneously. 

As detailed in \citeP24, the halo mass function and the cross-correlation functions of halos with different masses are extracted from the simulation and used to construct analytical fitting functions. We stress the fact that the framework introduced here is general, and can be used to predict the clustering and/or demographic properties of any populations of halo tracers (see also Appendix \ref{sec:details_modeling}).

We summarise below the main findings of our analysis: 
\begin{itemize}
    \item We jointly model all the observational data in Table \ref{tab:obs_data} except for the quasar auto-correlation function \citep[][]{arita2023}{}{}, which we analyze separately in Appendix \ref{sec:autos_arita}. We find a very good match between our predictions and observations for all the quantities considered (Fig. \ref{fig:overview}). The posterior distribution for the model parameters favors relatively large values for the scatter both in the quasar luminosity-halo mass relation and in the galaxy luminosity-halo mass relation ($\sigma\approx0.6-0.8\,\mathrm{dex}$), with the relation for quasars being steeper than the one for galaxies (Fig. \ref{fig:results_qg}, left panel). The active fraction, on the other hand, is larger for galaxies ($f_\mathrm{on}\approx25\%$) than for quasars ($\approx4\%$). Interestingly, the luminosity-halo mass relations inferred in Fig. \ref{fig:results_qg} (left) imply that galaxies outshine quasars (i.e, the average \OIII luminosity of galaxies is larger than the bolometric luminosity of quasars) at halo masses of $\log_{10} M/\msun \lesssim 11.5$.
    \item According to the results above, $z\approx6$ quasars live on average in $\approx10^{12.5}\,\msun$ halos, with a mass distribution that is quite broad, from $\approx10^{12.1}\,\msun$ halos to $\approx10^{12.8}\,\msun$ (according to the $2\sigma$ limits of the QHMF distribution; see right panel of Fig. \ref{fig:results_qg}). 
    This broad QHMF distribution implies that quasars inhabit rather diverse environments at high-$z$. This, together with the contribution of cosmic variance, may explain the large quasar-to-quasar variance in terms of environments that was reported by \citet[][]{eilers2024}{}{}, as well as the contradictory claims that have been made based on previous observations \citep[e.g.,][]{Kim09, Mazzucchelli2017, Mignoli20}{}{}. 
    \item Despite the rather large uncertainties, we are able to constrain the $z\approx6$ (UV-bright) quasar duty cycle to $\varepsilon_\mathrm{QDC}\lesssim1\%$ (Fig. \ref{fig:results_dc}, left panel). This relatively low value translates to quasar lifetimes of $\approx10\,\mathrm{Myr}$, in stark contrast with the very long lifetimes required at high $z$ by the standard picture of SMBH formation and growth \citep[e.g.,][]{inayoshi2020}{}{}.  This finding challenges our paradigm for SMBH growth at high-$z$, and suggests that most of the black hole mass growth may have happened in highly obscured and/or radiatively inefficient environments \citep[see also][]{Davies2019b}{}{}. 
    \item As expected, the properties of galaxies (i.e., \OIII emitters) that we obtain are rather different from the ones of quasars (Fig. \ref{fig:overview}-\ref{fig:results_dc}). The characteristic host mass for \OIII emitters that we measure from the GHMF is $\approx10^{10.9}\,\msun$, in line with the one estimated from LBG clustering measurements \citep[e.g.][]{barone-nugent2014,dalmasso2024}{}{}. This suggests that \OIII emitters may be tracing the population of high-$z$ actively star-forming galaxies in a way that is similar to what LBGs have been doing in the Hubble Space Telescope (HST) era. The galaxy duty cycle that we infer is larger than the one of quasars, $\varepsilon_\mathrm{GDC}\approx13\%$, suggesting that a significant fraction of high-$z$ galaxies are UV-bright and actively star-forming at $z\approx6$. This sets an implicit constraint on the fraction of galaxies that are quenched and/or obscured at the same redshifts.
    \item By comparing the properties of quasars at $z\approx6$ obtained in this work with the ones discussed in \citeP24 for $z\approx2.5$ and $z\approx4$, we find that the evolution of these properties with redshift seems to follow a non-monotonic trend (Fig. \ref{fig:results_z}). The characteristic quasar-host mass and the quasar duty cycle have similar values at $z\approx2.5$ and $z\approx6$, but they increase to significantly higher values at $z\approx4$ due to the strong quasar clustering measured by \citet[][]{shen2007}{}{}. We discuss whether the conjunction between $z\approx2.5$ and $z\approx6$ may suggest that quasar properties are more or less stable across cosmic time, which would imply that the $z\approx4$ quasar clustering measurements are overestimated. We also present a picture, however, in which the bulk of quasar activity takes place in very rare and over-dense environments (corresponding to $\approx5\sigma$ fluctuations in the initial linear density field) at $z\approx4$ and $z\approx6$, while it migrates to a larger population of less biased halos at lower-$z$. Further observational work is needed to distinguish between these scenarios and map the evolution of quasar properties across cosmic time.
\end{itemize}

The analysis presented in this paper lays down a simple but powerful framework that exploits observations to characterize the properties of SMBHs and galaxies in the early Universe. New data and more detailed modeling can improve the constraints that we get in the context of this framework significantly. 

Observationally, the ASPIRE survey \citep[][]{aspire_wang2023}{}{} will soon complement observations from EIGER \citep[][]{kashino2022, eilers2024}{}{} by measuring the cross-correlation function for a larger sample of $25$ moderately luminous quasars at $z\approx6.5-6.8$. The enlarged sample provided by ASPIRE will be extremely useful for reducing the uncertainties in our model parameters as well as for quantifying the quasar-to-quasar variance in the cross-correlation function. In the near future, new observations from JWST could complement the ASPIRE and EIGER surveys by determining the clustering properties of quasars and galaxies in a wider redshift range as well as for the faint end of the quasar luminosity function. 

In parallel with the acquisition of new observational data, the model presented here could be developed further to study the variance of the measured correlation function theoretically, and could be extended to take into account the velocity information coming from direct measurements of the redshift-space correlation function \citep[e.g.,][]{costa2023}{}{}. Results at different redshifts could also be linked together by developing an evolutionary model following the growth of supermassive black holes and the evolution of quasar activity across cosmic time.

\section*{Acknowledgements}

We are grateful to Junya Arita and the SHELLQs team for sharing their data on the quasar auto-correlation function and to Jan-Torge Schindler for discussion on the quasar luminosity function. 
We acknowledge helpful conversations with the ENIGMA group at
UC Santa Barbara and Leiden University. 
EP is grateful to Rob McGibbon and Victor Forouhar Moreno for help with the simulation outputs, and to Timo Kist, Jiamu Huang, and Vikram Khaire for comments on an early version of the manuscript.
JFH and EP acknowledge support from the European Research Council (ERC) under the European
Union’s Horizon 2020 research and innovation program (grant agreement No 885301).
This work is partly supported by funding from the European Union’s Horizon 2020 research and innovation programme under the Marie Skłodowska-Curie grant agreement No 860744 (BiD4BESt). 
FW acknowledges support from NSF grant AST-2308258. 
This work used the DiRAC Memory Intensive service (Cosma8) at the University of Durham, which is part of the STFC DiRAC HPC Facility (\url{www.dirac.ac.uk}). Access to DiRAC resources was granted through a Director’s Discretionary Time allocation in 2023/24, under the auspices of the UKRI-funded DiRAC Federation Project. The equipment was funded by BEIS capital funding via
STFC capital grants ST/K00042X/1, ST/P002293/1, ST/R002371/1 and ST/S002502/1,
Durham University and STFC operations grant ST/R000832/1. DiRAC is part of the
National e-Infrastructure.

\section*{Data Availability}
The derived data generated in this research will be shared on reasonable requests to the corresponding author.



\bibliographystyle{mnras}
\bibliography{biblio} 




\appendix

\section{Details on the conditional luminosity function framework} \label{sec:details_modeling}

Given any population of ``tracer'' (``T'') objects that are hosted by dark matter halos and are visible in some electromagnetic band, we can write down their 2-d distribution in the tracer
luminosity-host halo mass plane, $n( L, M)$, as:
\begin{equation}
   n(L, M) = {\rm CLF}(L| M) \,n_\mathrm{HMF}(M),
\end{equation}
where $n_\mathrm{HMF}(M)$ is the halo mass function. The quantity ${\rm CLF}(L| M)$ is known as the conditional luminosity function, and it links in a statistical sense the population of dark matter halos to the population of tracer objects \citep[e.g.,][]{yang2003, ballantyne2017a,ballantyne2017b,bhowmick2019, ren_trenti2020}. 

In this framework, we assume that every halo between a minimum mass $M_{\rm min}$ and a maximum mass $M_{\rm max}$ hosts a tracer object\footnote{$M_{\rm min}$ and $M_{\rm max}$ are chosen here according to the mass range that can be reliably modeled based on the cosmological simulation employed (see Sec. \ref{sec:simulations}).}. The luminosity $L$ of this tracer can be defined arbitrarily, but it has to depend solely on the mass of the halo. Following these assumptions, a simple marginalization of $n( L, M)$ over halo mass gives the luminosity function of the tracer species, $n_\mathrm{TLF}$:
\begin{align}
        n_\mathrm{TLF}( L) &= \int_{M_{\rm min}}^{M_{\rm max}} {\rm CLF}( L| M) \, n_\mathrm{HMF}( M)\,\d M . \label{eq:qlf}
\end{align}

Analogously, integrating over the luminosity dimension returns the distribution in mass of the tracers. If we include only objects above some threshold luminosity (set e.g. by the flux limit of observations), we can obtain a mass distribution for halos whose tracer object is brighter than $L_{\rm thr}$, $n_\mathrm{THMF}$:
\begin{equation}
    n_\mathrm{THMF}( M|L>L_{\rm thr}) = n_\mathrm{HMF}(M)\,\int_{ L_{\rm thr}}^\infty {\rm CLF}( L| M) \, \d  L .\label{eq:qhmf}
\end{equation}

Likewise, the aggregate probability for a halo of mass $M$ to host a tracer with a luminosity above $L_{\rm thr}$ (also known as a Halo Occupation Distribution, HOD; see e.g., \citealt{berlind2002}) is:
\begin{equation}
    {\rm HOD}(M) = \frac{n_\mathrm{THMF}( M|L>L_{\rm thr})}{n_\mathrm{HMF}(M)} = \int_{L_{\rm thr}}^\infty {\rm CLF}( L| M) \, \d  L .\label{eq:hod}
\end{equation}

Following, e.g., \citeP24 \citep[see also][]{ren_trenti2020}{}{}, we can define the duty cycle of tracers above the luminosity threshold, $\varepsilon_{\rm DC}$, as the weighted average of the HOD above a threshold mass that is given by the median of the tracer-host mass function, $n_\mathrm{THMF}( M|L>L_{\rm thr})$.
In other words, if we define the median of the $n_\mathrm{THMF}( M|L>L_{\rm thr})$ as the mass $M_{\rm med}$ satisfying the relation:
\begin{equation}
   \int_{M_{\rm med}}^{M_{\rm max} }n_{\rm THMF}(M) = 0.5\,\int_{M_{\rm med}}^{M_{\rm max}} n_{\rm THMF}(M), \label{eq:qhmf_med}
\end{equation}
then $\varepsilon_{\rm DC}$ can be expressed as:
\begin{equation}
\begin{split}
        \varepsilon_{\rm DC} &= \frac{\int_{M_{\rm med}}^{M_{\rm max}}  n_\mathrm{HMF}( M)\,{\rm HOD}(M) \,  \d M}{\int_{M_{\rm med}}^{M_{\rm max}} n_\mathrm{HMF}(M)\,\d M} =\\
        &= \frac{\int_{M_{\rm med}}^{M_{\rm max}} n_\mathrm{THMF}( M|L>L_{\rm thr})\,\d M}{\int_{M_{\rm med}}^{M_{\rm max}} n_\mathrm{HMF}( M)\,\d M}.\\
        \label{eq:duty_cycle}
        \end{split}
\end{equation}

These relations hold for any tracer populations that satisfy the assumptions made above. In \citeP24, we have considered SMBHs as tracer objects, assuming that every halo hosts a SMBH at its center emitting at some luminosity $L$. If the luminosity $L$ is high enough, the SMBH becomes an active quasar, and so we can use the conditional luminosity framework to obtain the quasar luminosity function ($n_\mathrm{QLF}$; analogous to eq. \ref{eq:qlf}), the quasar-host mass function ($n_\mathrm{QHMF}$; analogous to eq. \ref{eq:qhmf}), and the quasar duty cycle ($\varepsilon_{\rm QDC}$; analogous to \ref{eq:duty_cycle}).

As commonly assumed in the literature \citep[e.g.,][]{yang2003,vandenbosch2003}{}{}, galaxies are also tracers of the dark matter halo distribution. Following the \citeP24 approach, we can then assume a conditional luminosity function for galaxies, and adapt the relations above to obtain the galaxy luminosity function ($n_\mathrm{GLF}$; analogous to eq. \ref{eq:qlf}), the galaxy-host mass function ($n_\mathrm{GHMF}$; analogous to eq. \ref{eq:qhmf}), and the galaxy duty cycle  ($\varepsilon_{\rm GDC}$; analogous to \ref{eq:duty_cycle}).  

In Sec. \ref{sec:population_model}, we write down explicitly the quasar/galaxy conditional luminosity functions adopted in this work\footnote{Note that these functions depend on the specific population of ``quasars'' and ``galaxies'' we model, as well as on the definition of their luminosity, $L$. We refer to Sec. \ref{sec:population_model} for more details.}, and provide more details on how to connect the quantities defined here to observations.

\section{Results for the fitting of the halo cross-correlation functions} \label{sec:details_fitting}

As described in Sec. \ref{sec:simulations}, we compute the cross-correlation functions between $z\approx6$ halos in different mass bins, $\xi_h(M_j,M_k;r)$, and then fit the results with a suitable parametrization of the radial and mass dependences. The details of the fitting are summarized in the main text and described at length in \citeP24. Here, we focus on the results of these fits, comparing them to the actual correlation functions computed numerically from simulations and discussing their validity in the context of the problem we are facing here. 

Figure \ref{fig:simulations_fit_triangle} displays the overall results of the fit. The first two rows display the resulting fitting function ($\rho_\mathrm{fit}(\nu(M_j), \nu(M_k), r)=\xi_h(M_j,M_k;r)/\xi_\mathrm{ref}(r)$, where $\xi_\mathrm{ref}(r)$ is a reference correlation function, see main text for details). Each panel in these rows show the values of $\rho_\mathrm{fit}(\nu(M_j), \nu(M_k), \Bar{r})$ as a function of the two masses $M_j$ and $M_k$ at a different scale $\Bar{r}$. The last two rows show the relative difference ($\rho/\rho_\mathrm{fit} -1$) between our fit and the values of $\rho(\nu(M_j), \nu(M_k), r)=\xi_h(M_j,M_k;r)/\xi_\mathrm{ref}(r)$ obtained from the simulation. According to these figures, our simple analytical framework can describe the behavior of cross-correlation functions for a wide range of masses and scales with a good degree of accuracy ($\lesssim 5-10\%$). This level of accuracy is sufficient for the data we aim to reproduce here, as both the auto-correlation function of quasars and the quasar-galaxy cross-correlation functions are only known at the $30\%-100\%$ level. The most constrained quantity is the galaxy auto-correlation function, which is however still uncertain at more than $\gtrsim10\%$ (Sec. \ref{sec:observations}).

The only notable exception for which our fit doesn't perform well is the case of high masses ($\log_{10} M_{j,k}/\msun \approx11.5$) and small scales ($r\lesssim0.5\,\cMpc$). However, this behavior is expected as high-mass halos are quite rare, and hence the measured correlation functions suffer in general from significant shot noise. At small scales this is worsened by the fact that the correlation function is dominated by the clustering of satellite halos, which are in general less massive than $\log_{10} M /\msun \approx11-12$. As a result, the cross-correlation functions of very massive systems drop at $r\lesssim0.5\,\cMpc$ because of halo exclusion. Our fit hinges upon a smooth dependence of the correlation functions on mass and radius, and it is not able to capture halo exclusion properly. Nonetheless, this is not an issue for our analysis, because the data we aim to fit do not probe this specific regime: the auto-correlation function of quasars from \citet[][]{arita2023}{}{} is only measured at very large scales ($r\gtrsim40\,\cMpc$), while the quasar-galaxy cross-correlation function and the auto-correlation function of galaxies from \citeE24 are dominated by the contribution of galaxies, which live in relatively low mass halos ($\log_{10} M /\msun \approx10.5-11$; see Section \ref{sec:results}).

Figure \ref{fig:simulations_fit} shows two more comparisons between the cross-correlation functions extracted from the simulation and our fitting functions. In the left panel, we show the cross-correlation terms $\xi_h(M,\Tilde{M},r)$ as a function of radius, for different values of the mass $M$. The mass $\Tilde{M}$ is chosen to represent the bin $\log_{10} M /\msun=10.5-10.75$. Errors on the values extracted from simulations are Poissonian (Sec. \ref{sec:simulations}). Note that to properly reproduce the correlations measured in simulations, we select halos in each mass bin and weigh the fitting functions according to the mass distribution of halos (i.e., the halo mass function). In this way, we can take into account the actual distribution of halo masses in our fitting framework.
Overall, we confirm that the fits and the values from simulations agree at the $\approx5-10\%$ level, with the expected exception of the most inner bin. 

The right panel of Figure \ref{fig:simulations_fit} shows the halo auto-correlation functions for each mass bin, $\xi_h(M,M,r)$. As already mentioned above, we note that the accordance between fits and simulations is again satisfactory with the notable exceptions of large halo masses -- for which halos are rare and the measured correlation functions are noisy -- and small scales -- for which halo exclusion plays an important role and our fit is not able to capture it properly. Overall, this visual comparison between simulations and fits confirms the fact that our framework can properly reproduce cross-correlation functions at all scales, as well as auto-correlation functions, with the exception of the high mass bins at small scales.

 \begin{figure*}
	\centering
	\includegraphics[width=\textwidth]{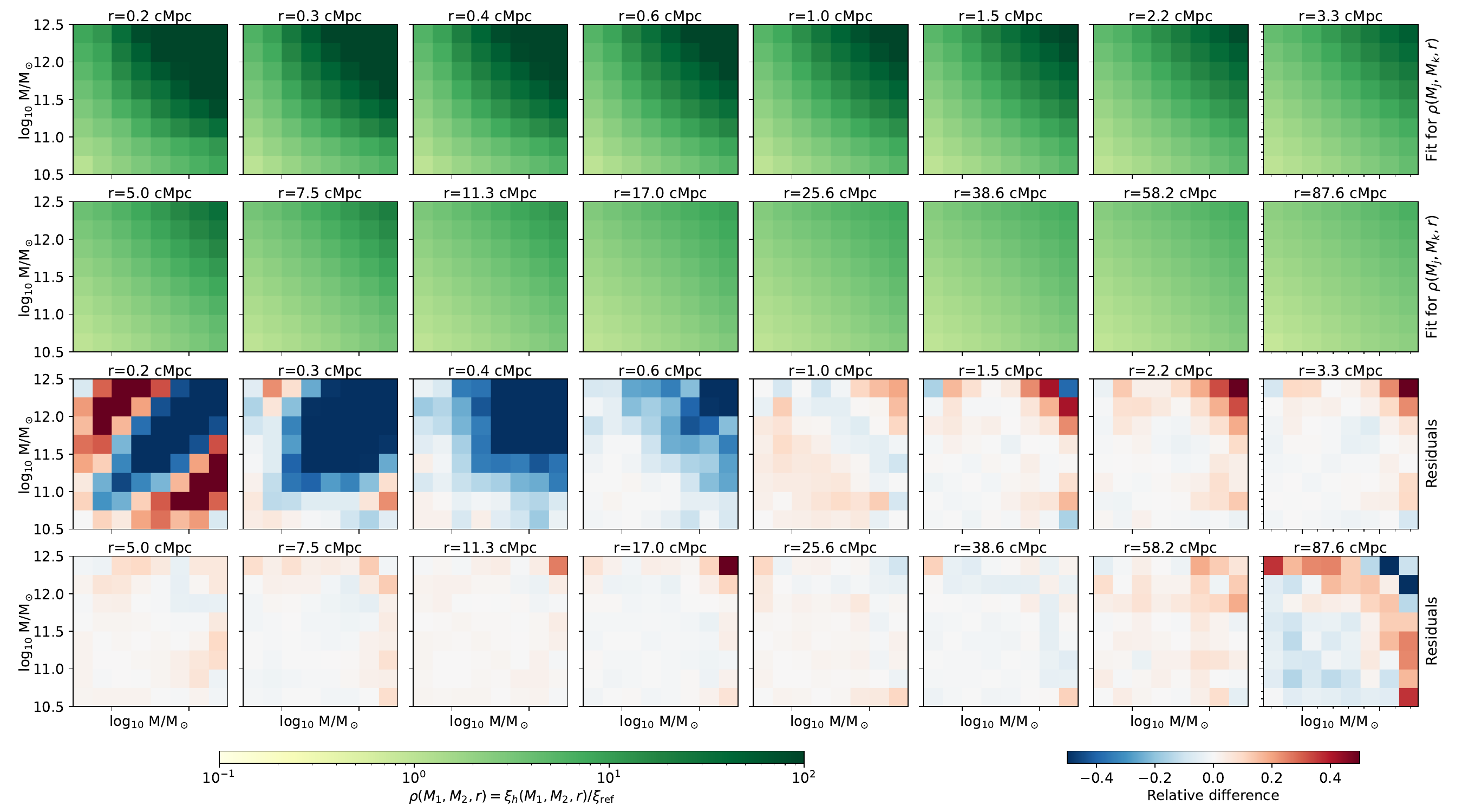}

	 \caption{Results for the fitting of the $z\approx6$ cross-correlation terms $\rho(M_j, M_k, r)=\xi_h(M_j,M_k;r)/\xi_\mathrm{ref}(r)$ (see Appedix \ref{sec:details_fitting} for definitions). The two top rows show the fitting function $\rho_\mathrm{fit}(M_j, M_k, r)$ as a function of the two masses $M_1$ and $M_2$ for different values of the distance $r$. The last two rows show the relative difference between the fits and the values extracted from simulation.   
  \label{fig:simulations_fit_triangle}
 	}
\end{figure*}

 \begin{figure*}
	\centering
	\includegraphics[width=\textwidth]{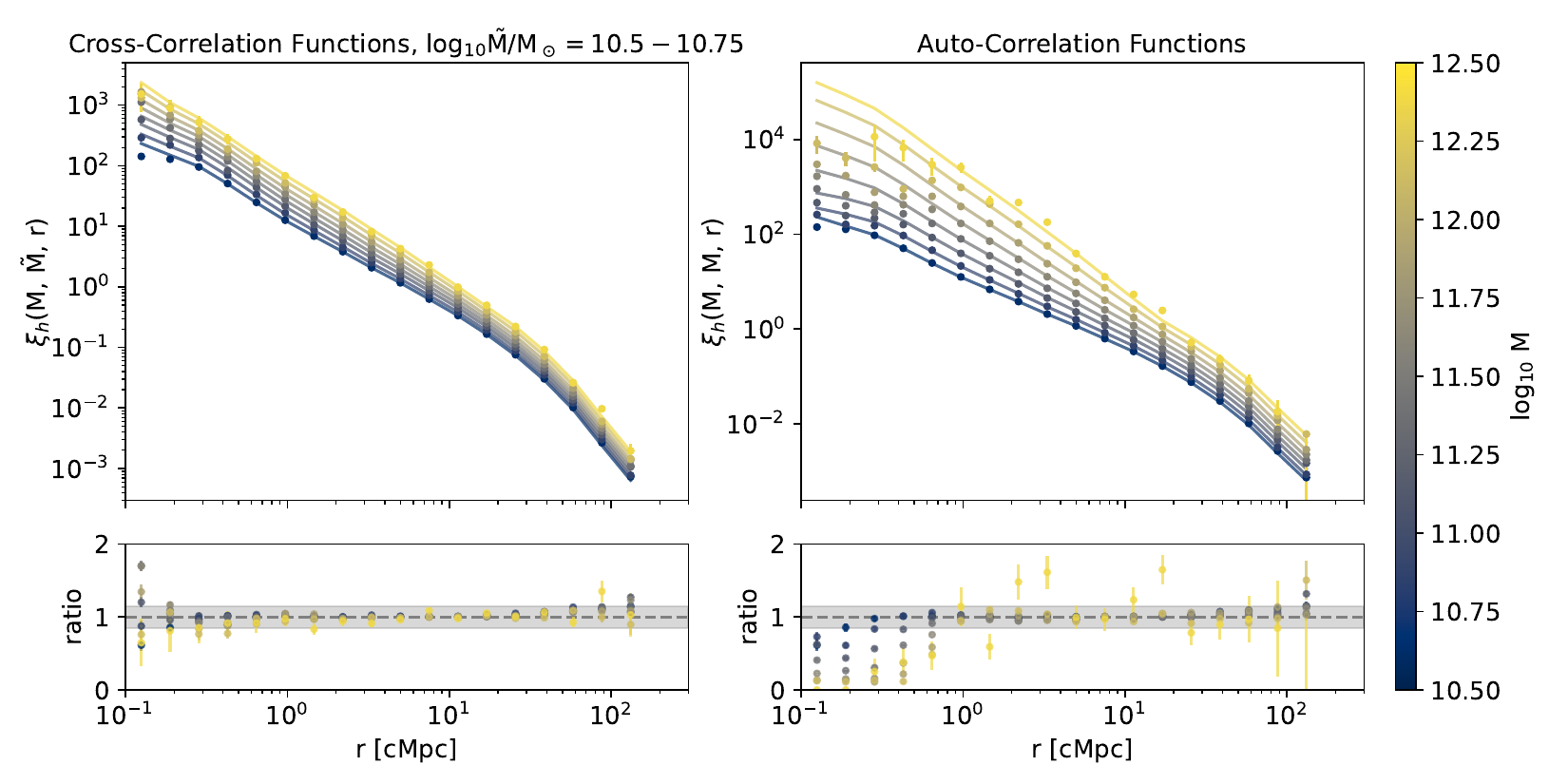}

	 \caption{{\it Left:} Cross-correlation functions of halos in different mass bins, $\xi_h(M,\Tilde{M},r)$, at $z\approx6$. The mass $\Tilde{M}$ is set to correspond to the $\log_{10} M/\msun = 10.5-10.75$ bin, while the other mass is varied according to the color scale. Values extracted from simulations are shown as data points, with error bars given by the Poissonian statistics of pair counting (see Sec. \ref{sec:simulations}). Solid lines represent the fitting functions to these simulated values. Relative differences between the fit and the simulation are shown in the bottom panel. {\it Right:} Same as the left panel, but for the auto-correlation functions of halos in different mass bins, $\xi_h(M,M,r)$.  \label{fig:simulations_fit}
 	}
\end{figure*}

\section{Interpreting the auto-correlation measurements of $z\approx6$ quasars}
\label{sec:autos_arita}

In this Section, we analyze the data concerning the quasar auto-correlation function from \citet{arita2023}. As detailed in Sec. \ref{sec:observations}, we decided to leave this dataset out of the joint fit performed in the main analysis because we realized that its constraining power was less strong than expected. In particular, we found that using only the \citet[][]{arita2023}{}{} data, we were not able to place significant constraints on any of our model parameters. 

For this reason, we use here a much simpler model that should make the interpretation of the data straightforward. In particular, we choose to parameterize the quasar-host mass function (QHMF) in the following way: 
\begin{equation}
    n_{\rm QHMF}(M)=\varepsilon_{\rm DC}n_{\rm HMF}(M)\Theta(\log_{10} M - \log_{10} M_{\rm min}), \label{eq:qhmf_arita}
\end{equation}
with $\varepsilon_{\rm QDC}$ being the duty cycle and $\Theta$ the Heaviside step function. 
In practice, we assume a simple ``step-function'' halo occupation distribution (HOD) model, depending only on one single parameter, the minimum host mass, $M_{\rm min}$ (the duty cycle $\varepsilon_{\rm QDC}$ is completely irrelevant for clustering measurements). 

For every value of $M_{\rm min}$, we can take the resulting QHMF and use it to compute the quasar auto-correlation function, $\xi_\mathrm{QQ}(r)$, according to eq. \ref{eq:quasar_corr_func}. With a simple integration along the radial direction (eq. \ref{eq:projected_corrfunc}), we can then obtain the projected auto-correlation function, $w_{p,\mathrm{QQ}(r_p)}$, which can be compared directly with the \citet[][]{arita2023}{}{} data. 

As detailed in Sec. \ref{sec:dmo_fit}, our model for the correlation functions consists of two components: a fit to simulations, $\xi_{h,\mathrm{fit}}$, and a prediction based on the linear halo bias formalism, $\xi_{h,\mathrm{lin}}$ (eq. \ref{eq:corr_linear}). The former is used to model the small-scale clustering ($r\lesssim20\,\cMpc$), while the latter is used to regularize the behaviour of simulations at large scales ($r\gtrsim20\,\cMpc$). The key point, here, is that the \citet[][]{arita2023}{}{} data we aim to interpret cover only very large scales, with the innermost bin at $r\approx40\,\cMpc$. For this reason, we can safely assume that our model is entirely in the linear theory regime, and assume $\xi_h = \xi_{h,\mathrm{lin}}$. In other words, the model we discuss in this context is not unique to our simulations; instead, it is very general and solely based on the linear growth of structures in a $\Lambda$CDM cosmology.

The left panel of Fig. \ref{fig:autos_arita} shows the predictions for the projected correlation function according to our ``linear theory'' model, for different values of the minimum host mass $M_{\rm min}$. These are compared with data in a quantitative way by determining the $\chi^2$ statistics for each $M_{\rm min}$ in the left panel of Fig. \ref{fig:autos_arita}. The $\chi^2$ is computed by taking into account the covariances between different data points. We see that we obtain values of the $\chi^2$ in the range $\chi^2\approx6-7$, which are perfectly compatible with data and translate into reduced chi-squared values of $\approx1.5-1.75$. There is a slight preference in our model for smaller values of the minimum host mass, but the measurement is not statistically significant for any reasonable values of $\log_{10} M_\mathrm{min}/\msun\lesssim13.5$.

The conclusion obtained here in the context of our model differs from the one found by \citet[][]{arita2023}, who analyzed the same data and measured a rather high value of the characteristic host halo mass for quasars at $z\approx6$, i.e., $\log_{10} M/M_\odot=12.9^{+0.4}_{-0.7}$. 
The striking difference between our conclusions and the ones in the \citet[][]{arita2023}{}{} analysis may reside in the different assumptions made for the shape and normalization of the correlation functions. While we assume physically-motivated halo correlation functions that follow linear theory, and convert these into a quasar-correlation function in a second step, \citet[][]{arita2023}{}{} parametrize the quasar auto-correlation function directly by assuming a power-law shape with a slope of $-1.8$ and a normalization set by the quasar auto-correlation length, $r_{0,\mathrm{QQ}}$. The results for this parametrization are also shown in Fig. \ref{fig:autos_arita} with green shadings (with the corresponding chi-squared values shown in the right panel).   It is quite interesting to see that the power-law shaped models for the quasar auto-correlation functions reach a better agreement with the data than the linear theory ones, with a minimum $\chi^2\lesssim5$ corresponding to large values of the auto-correlation length  ($r_{0,\mathrm{QQ}}\approx20-50\,\cMpc$), in agreement with the findings of \citet[][]{arita2023}{}{}. 

We conclude by noting that our model presented in the main analysis (Sec. \ref{sec:results}) is compatible with the data from \citet[][]{arita2023}{}{}. Indeed, if we take the best-fit parameters from Fig. \ref{fig:overview} and compare the prediction for the quasar auto-correlation function with data we find a value for the chi-square of $\chi^2\approx6$, which is consistent with the discussion above and implies a good match with observations. This implies that the \citet[][]{arita2023}{}{} measurements are perfectly compatible with the clustering constraints from JWST (\citeE24). However, the \citet[][]{arita2023}{}{} data are very uncertain and limited only to very large scales. As a consequence, they result in rather weak constraints that -- as shown in this Section -- are very sensitive to the exact prescription made for the shape of the quasar auto-correlation function.

 \begin{figure*}
	\centering
	\includegraphics[width=\textwidth]{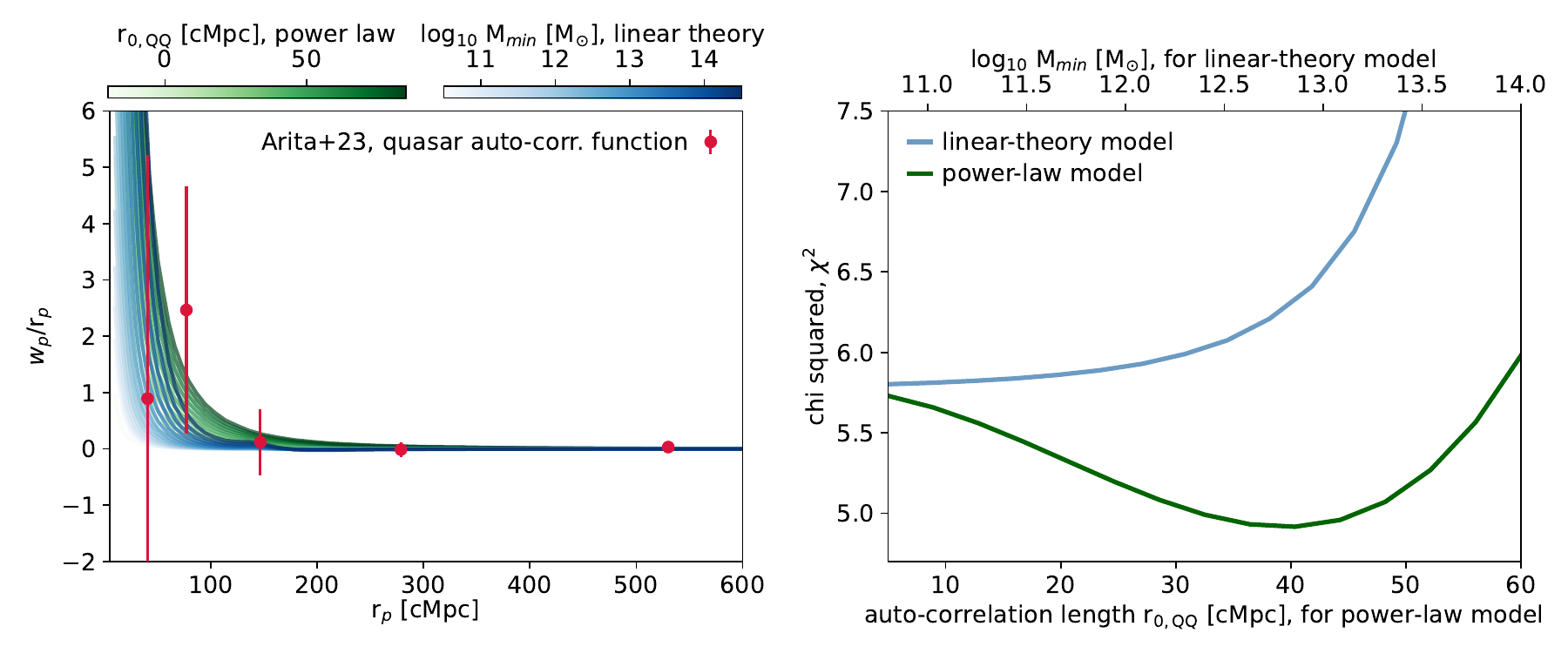}

	 \caption{\textit{Left}: Projected $z\approx6$ quasar auto-correlation function, $w_p/r_p$ as a function of the distance, $r_p$. The observational data from \citet[][]{arita2023}{}{} are shown as red points. Predictions from the model based on linear theory, which is our fiducial one, are shown as blue lines, color-coded based on the value of the $M_\mathrm{min}$ parameter (eq. \ref{eq:qhmf_arita}). Predictions coming from a power-law model for the correlation functions are shown in green, color-coded according to the value of the quasar auto-correlation length, $r_{0,\mathrm{QQ}}$. 
    \textit{Right}: Comparison of model predictions with data, according to the value of the $\chi^2$ statistic. The blue line refers to the ``linear theory'' model, and it is parametrized by the minimum host mass $M_\mathrm{min}$ (top label). The green line, instead, refers to the ``power-law'' model and is parametrized by the quasar auto-correlation length, $r_{0,\mathrm{QQ}}$ (bottom label).
  \label{fig:autos_arita}
 	}
\end{figure*}

\section{Quasar-host halo masses with a uniform luminosity threshold} \label{sec:same_Lthr}

As discussed in Sec. \ref{sec:discussion_redshift}, the quasar host mass functions (QHMFs) shown in Fig. \ref{fig:results_z} are obtained by setting a luminosity threshold for modeling quasar clustering that varies with redshift according to the one employed in observations. Here, we show (Fig. \ref{fig:results_z_Lthr}) the effect of setting a uniform luminosity threshold of $\log_{10} L_\mathrm{thr}/\ergs =46.7$ at all redshifts. This threshold corresponds to the one employed at $z\approx4$, so the $z\approx4$ results are the same as in Fig. \ref{fig:results_z}. The QHMF at $z\approx2$ ($z\approx6$) shifts to higher (lower) masses respectively, due to the different quasar population probed by the \citet[][]{eftekharzadeh2015}{}{} (\citeE24) data. This effect, however, is not strong enough to impact in any relevant way the discussion on the evolution of quasar properties with redshift made in Sec. \ref{sec:discussion_redshift}.

\begin{figure*}
	\centering
	\includegraphics[width=\textwidth]{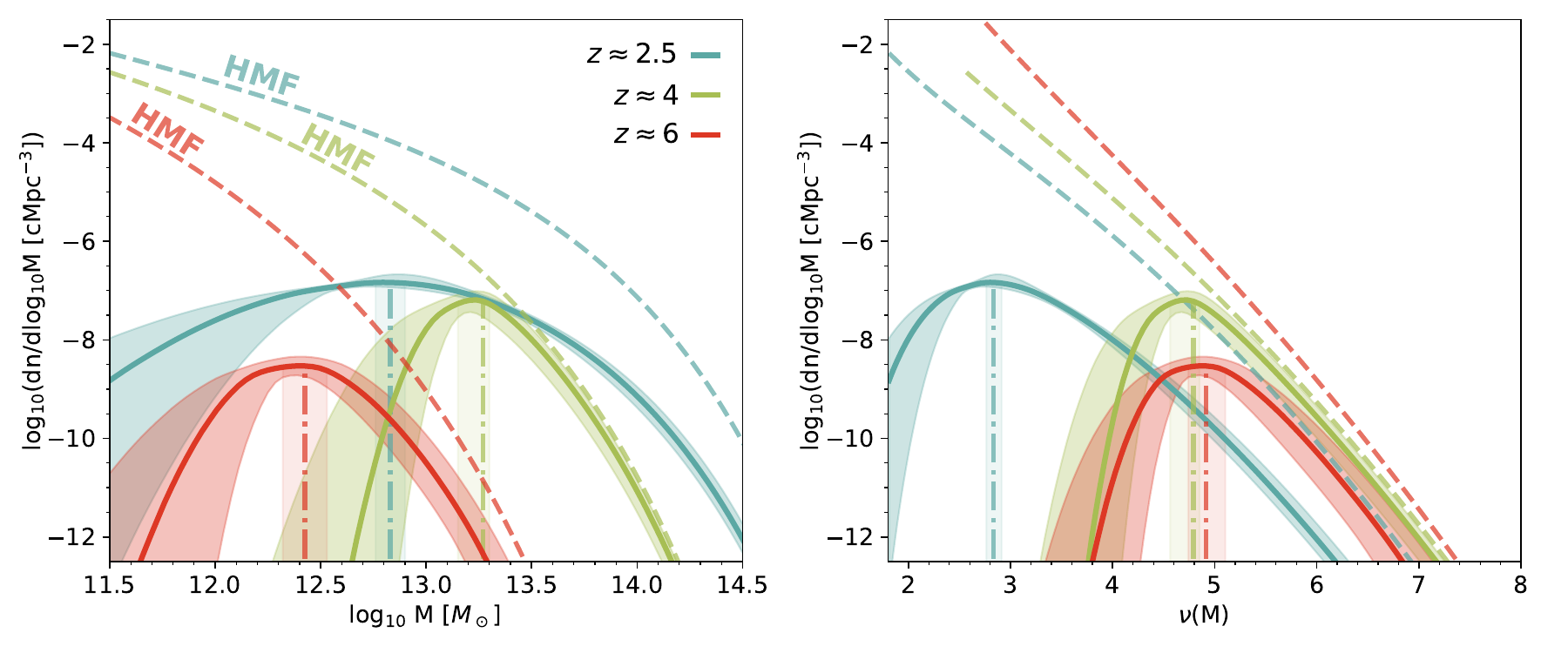}

	 \caption{ Same as Fig. \ref{fig:results_z}, but the QHMFs here are obtained by setting a uniform luminosity threshold for the clustering measurements at all redshifts, i.e., $\log_{10} L_\mathrm{thr}/\ergs =46.7$. The QHMF represents the mass distribution of halos that are hosting quasars brighter than $L_\mathrm{thr}$.
  \label{fig:results_z_Lthr}
 	}
\end{figure*}


\bsp	
\label{lastpage}
\end{document}

%% file: additional_tex/pack.tex
\usepackage{newtxtext,newtxmath}

\usepackage[T1]{fontenc}
\usepackage{ae,aecompl}
\usepackage[utf8]{inputenc}

\usepackage{graphicx}	

\usepackage{amsmath}	
\usepackage{amssymb}	
\usepackage{comment}
\usepackage{siunitx}
\usepackage{booktabs}
\usepackage{gensymb}
\usepackage{threeparttable} 

\usepackage{xcolor}

%% file: additional_tex/definitions.tex
\renewcommand{\d}{\mathrm{d}}

\let\oldnabla\nabla
\renewcommand{\nabla}{\vec{\oldnabla}}

\def\be{\begin{equation}}
\def\ee{\end{equation}}
\newcommand\code[1]{\textsc{\MakeLowercase{#1}}}

\def\gsim{\lower.5ex\hbox{\gtsima}} 
\def\lsim{\lower.5ex\hbox{\ltsima}} 
\def\gtsima{$\; \buildrel > \over \sim \;$} 
\def\ltsima{$\; \buildrel < \over \sim \;$} \def\gsim{\lower.5ex\hbox{\gtsima}} 
\def\lsim{\lower.5ex\hbox{\ltsima}} 
\def\simgt{\lower.5ex\hbox{\gtsima}} 
\def\simlt{\lower.5ex\hbox{\ltsima}}

\def\msun{{\rm M}_{\odot}}

\def\kms{\,\rm km\,s^{-1}}

\def\ergs{{\rm erg}\,{\rm s}^{-1}}

\def\S*{$\Sigma_{\rm SFR}$}

\def\OIII{\hbox{[O~$\scriptstyle\rm III $]~}}

\def\cMpc{{\rm cMpc}}
\def\cGpc{{\rm cGpc}}
\def\cMpch{{\rm cMpc} \,{\rm h}^{-1}}